\documentclass[12pt]{article}

\usepackage{amssymb}
\usepackage{amsmath}
\usepackage{amscd}
\usepackage{latexsym}
\usepackage{graphicx}

\usepackage{cite}

\newcommand{\be}{\begin{equation}}
\newcommand{\ee}{\end{equation}}
\newcommand{\Dlt}{\Delta}
\newcommand{\dlt}{\delta}
\newcommand{\prt}{\partial}
\newcommand{\br}{{\bf r}}
\newcommand{\bk}{{\bf k}}

\newcommand{\bt}{\beta}
\newcommand{\vp}{\varphi}
\newcommand{\ep}{\varepsilon}
\newcommand{\al}{\alpha}
\newcommand{\ra}{\rightarrow}
\newcommand{\sgm}{\sigma}

\newcommand{\gm}{\gamma}
\newcommand{\om}{\omega}
\newcommand{\Om}{\Omega}
\newcommand{\Gm}{\Gamma}
\newcommand{\dgr}{\dagger}
\newcommand{\lbd}{\lambda}
\newcommand{\Lbd}{\Lambda}

\newcommand{\cL}{{\cal L}}
\newcommand{\rgl}{\rangle}
\newcommand{\lgl}{\langle}

\begin{document}

\begin{center}
{\Large {\bf Major issues in theory of Bose-Einstein condensation
} \\ [5mm]

V.I. Yukalov$^{1,2}$ } \\ [3mm]

{\it $^1$Bogolubov Laboratory of Theoretical Physics, \\
Joint Institute for Nuclear Research, Dubna 141980, Russia \\ [2mm]

$^2$Instituto de Fisica de S\~ao Carlos, Universidade de S\~ao Paulo, \\
CP 369, S\~ao Carlos 13560-970, S\~ao Paulo, Brazil} \\ [3mm]

{\bf E-mail}: yukalov@theor.jinr.ru

\end{center}

\vskip 2cm

\begin{abstract}
Major issues arising in the theory of Bose-Einstein condensation are reviewed. These
issues, although being principally important, are very often misunderstood, which
results in wrong conclusions. The basic point is global gauge symmetry breaking that
is a necessary and sufficient condition for Bose-Einstein condensation. Paying no
attention to this basic point is a common fallacy leading to a number of confusions.
For instance, the attempt of describing Bose condensation without gauge symmetry
breaking produces the so-called ``grand canonical catastrophe" that actually does not
exist in the correct description of Bose condensation accompanied by gauge symmetry
breaking. The other common flaw is forgetting to consider the stability of the studied
systems. One sometimes accomplishes lengthy calculations and discusses the properties of
a system that in reality cannot exist being unstable. In some cases, the seeming
instability is caused by the negligence of the simple mathematical reason teaching us
that one should not go beyond the approximation applicability. An example of such an
artificial instability is related to the appearance of the so-called ``thermodynamically
anomalous fluctuations" whose arising is due to the use of a second-order approximation
for calculating fourth-order terms, in this way distorting the $O(2)$-class model of
a Bose-condensed system to the Gaussian-class model. These and other principal points,
important for the correct treatment of Bose-condensed systems, are reviewed, including the
resolution of the Hohenberg-Martin dilemma of gapless versus conserving theories for
Bose-condensed systems and the problem of statistical ensemble equivalence.
\end{abstract}

\newpage

\section*{Content}

{\parindent=0pt
{\bf I}. Introduction

\vskip 3mm
{\bf II}. Criterion of Bose-Einstein condensation

\vskip 3mm
{\bf III}. Gauge symmetry breaking
\vskip 2mm
\hspace{1.5cm}   A. Gauge symmetry
\vskip 2mm
\hspace{1.5cm}   B. Method of quasi-averages
\vskip 2mm
\hspace{1.5cm}   C. Bogolubov theorem
\vskip 2mm
\hspace{1.5cm}   D. Condensate existence
\vskip 2mm
\hspace{1.5cm}   E. Ginibre theorem
\vskip 2mm
\hspace{1.5cm}   F. Roepstorff theorem
\vskip 2mm
\hspace{1.5cm}   G. Bogolubov shift

\vskip 3mm
{\bf IV}. Unitary inequivalent representations

\vskip 3mm
{\bf V}. Thermodynamic stability

\vskip 3mm
{\bf VI}. Dynamic stability

\vskip 3mm
{\bf VII}. Absence of ``grand canonical catastrophe"

\vskip 3mm
{\bf VIII}. Instability of ideal Bose gas

\vskip 3mm
{\bf IX}. Role of dimensionality

\vskip 3mm
{\bf X}. Power-law trapping potentials

\vskip 3mm
{\bf XI}. Resolution of Hohenberg-Martin dilemma

\vskip 3mm
{\bf XII}. Condensate wave function

\vskip 3mm
{\bf XIII}. Self-consistent mean-field approach

\vskip 3mm
{\bf XIV}. Normal versus symmetry-broken averages

\vskip 3mm
{\bf XV}. Absence of thermodynamically anomalous fluctuations

\vskip 3mm
{\bf XVI}. Fluctuation indices for composite observables

\vskip 3mm
{\bf XVII}. Equivalence of statistical ensembles

\vskip 3mm
{\bf XVIII}. Conclusion

}

\newpage

\section{Introduction}

Bose-Einstein statistics and the related Bose-Einstein condensation (BEC) was
advanced 100 years ago \cite{Bose_1,Einstein_2,Einstein_3}. The microscopic theory
of weakly interacting Bose gases at low temperature was put forward by Bogolubov
\cite{Bogolubov_4} in 1947 (see also the books \cite{Bogolubov_5,Bogolubov_6,Bogolubov_7}).
The Bogolubov theory proved to well agree with the experiments on BEC in very weakly
interacting trapped Bose gases at low temperature, first observed in 1995
\cite{Anderson_8,Bradley_9,Davis_10}.

However, in the theory of BEC at finite temperature and interactions till nowadays there
remain several problems not well understood and resulting in a number of confusions
that persuasively appear in many publications. This is the aim of the present review
to discuss the major subtle points of the BEC theory at arbitrary interactions and
temperature.

The criterion for the occurrence of BEC is recalled in Sec. II. The principal starting
point of the theoretical description of BEC is the understanding that for the very
existence of BEC the global gauge symmetry breaking is necessary and sufficient, which
is a rigorous mathematical fact (Sec. III). Forgetting this fact leads to various
fallacies and wrong conclusions, as will be shown below. From the mathematical point
of view, the systems with broken gauge symmetry and without its breaking live in
different Fock spaces that are orthogonal to each other, so that, when BEC occurs, it
is basically wrong to continue working in a gauge symmetric Fock space (Sec. IV).

The other point, that is often forgotten, is the necessity of checking the stability
of the studied system. Formally, in theory, the system may show BEC, although, being
unstable, it actually cannot exist, which happens with the uniform ideal Bose gas in
three dimensions. Sometimes, one distinguishes thermodynamic stability (Sec. V) and
dynamic stability (Sec. VI), which for an equilibrium system result in the same stability
conditions.

In the current literature, to great surprise, there is a very widely spread myth on the
existence of the so-called ``great canonical catastrophe", when particle fluctuations
of BEC become catastrophically large. One even tries to observe them in experiment,
forgetting that such catastrophic fluctuations would make the system extremely unstable,
hence it would not be able to exist at al. Meanwhile, there is no any ``great canonical
catastrophe", but its appearance is caused by the incorrect treatment of BEC without
breaking gauge symmetry. In a system with broken gauge symmetry, no catastrophic
condensate fluctuations arise, but moreover the condensate particle fluctuations are
actually absent (Sec. VII).

The catastrophic condensate fluctuations exist neither in an ideal gas nor in an
interacting Bose gas. Nevertheless, in a uniform ideal Bose gas in three dimensions,
there are large fluctuations of non-condensed particles, which implies the ideal-gas
instability (Sec. VII). Fortunately, any infinitesimally small interactions stabilize
the gas.

Stability of Bose-condensed systems depends on the space dimensionality. Even ideal
uniform Bose-condensed gas can be stabilized in dimensionality $d > 4$ (Sec. IX). 
Stability of trapped Bose gases depends on the trap shape, which is shown for power-law 
trapping potentials (Sec. X).

The description of realistic Bose-condensed systems of interacting atoms confronts the
Hohenberg-Martin dilemma of conserving versus gapless theories. The problem is that, as
soon as BEC occurs, hence global gauge symmetry becomes broken, the standard treatment
of BEC leads either to the appearance of an unphysical gap in the spectrum of collective
excitations or to the inconsistency of thermodynamic relations. The resolution of this
dilemma requires the use of a representative ensemble, where all conditions uniquely
characterizing the system are to be taken into account. Then we get a self-consistent
theory that is conserving and gapless (Sec. XI).

The difference between the condensate wave function and coherent vacuum field is
elucidated (Sec. XII). The vacuum field, whose mathematical structure is that of the
nonlinear Schr\"{o}dinger equation, does not exhaust the conventional mean-field picture,
but it is just a very particular case that can be applied to weakly interacting Bose gases
at zero temperature. The overall mean-field approach is presented by the
Hartree-Fock-Bogolubov approximation (Sec. XIII). The latter, being employed in the frame
of a representative ensemble, is self-consistent, conserving, and gapless, at the same
time taking account of gauge symmetry breaking. While the Hartree and Hartree-Fock
approximations, not respecting gauge symmetry breaking, cannot be used for characterizing
Bose-condensed systems.

As soon as global gauge symmetry is broken, the theory has to take into account both the
normal as well as symmetry-broken averages. The latter are termed as anomalous averages,
although being not merely regular for a system with broken gauge symmetry, but moreover
being compulsory for its correct description (Sec. XIV). The omission of the symmetry-broken
averages makes the theory not self-consistent, quantitatively incorrect, and acquiring
several nonphysical divergences.

The other important warning is the necessity of being consistently in the frame of a
chosen approximation, since going beyond the approximation applicability can lead to
wrong conclusions possessing no physical meaning. A typical example of such an arising
inconsistency is the appearance of the so-called ``thermodynamically anomalous
fluctuations" that has been a hot topic in recent years. However such ``anomalous
fluctuations", inducing system instability, are nothing but an artefact caused by
calculating fourth-order expressions in a second-order approximation, thus reducing
the $O(2)$-class model to a Gaussian-class model (Sec. XV).

In Sec. XVI, it is explained that thermodynamically anomalous fluctuations of composite
observables could arise if and only if at least one of the terms exhibits such anomalous
fluctuations. The problem of statistical ensemble equivalence is discussed in Sec. XVII,
where it is explained that the majority of claims of ensemble nonequivalence are caused
by confusions. Section XVIII summarizes the main points that have to be followed for
getting a self-consistent theory of systems with Bose-Einstein condensate, valid at
arbitrary interactions and temperature.

As far as the review is theoretical, there will be the need in writing some formulas. Of
course, intermediate calculations will be omitted, but the basic mathematical expressions
will have to be mentioned in order to avoid ambiguity. Throughout the paper, the system
of units is accepted where the Planck constant $\hbar$ and Boltzmann constant $k_B$ are
set to one.

\section{Criterion of Bose-Einstein condensation}

Talking about Bose-Einstein condensation (BEC), the first thing that has to be recalled
is the criterion of Bose-Einstein condensation. In simple words Bose-Einstein condensation
implies a macroscopic occupation of a quantum state. But one has to specify how the
quantum states are to be defined. There are qualitative signs showing when the phenomenon
of BEC is possible, which tell us that this happens when the thermal wavelength
becomes much larger than the mean interatomic distance. However, we need a general
criterion expressed in exact mathematical terms. Probably, the most general mathematical
formulation of BEC is due to Penrose and Onsager \cite{Penrose_11,Penrose_12}, which
is recalled below.

Statistical properties of a system are contained in correlation functions for the
operators describing the system. The basic correlation functions are reduced density
matrices \cite{Davidson_13,Coleman_14}. For simplicity, below we consider spinless
particles. Spins and other internal degrees of freedom can be incorporated by
representing the field operators as columns with respect to these degrees of freedom.
The first-order density matrix
\be
\label{1}
\rho(\br,\br')  \; = \; {\rm Tr} \; \psi(\br) \; \hat\rho \; \psi^\dgr(\br') \; = \;
\lgl \;  \psi^\dgr(\br') \;  \psi(\br) \; \rgl
\ee
is expressed through the field operators $\psi$ and a statistical operator ${\hat \rho}$. 
The field operators depend on the spatial variables ${\bf r}$ and on time $t$ that, for
brevity, is not shown explicitly. The eigenproblem
\be
\label{2}
\int \rho(\br,\br')  \; \vp_k(\br') \; d\br' \; = \; n_k \; \vp_k(\br)
\ee
defines the eigenfunctions, called natural orbitals, and, respectively, eigenvalues. The 
latter are named occupation numbers showing the number of particles occupying quantum states
labelled by a multi-index $k$. For uniform systems, $k$ can be a momentum, but generally
it is an index appropriate for distinguishing quantum states and natural orbitals.

The maximal of the occupation numbers
\be
\label{3}
 n_k \; = \; \int \vp_k^*(\br) \; \rho(\br,\br') \;
\vp_k(\br') \; d\br d\br' \;  ,
\ee
in equilibrium corresponding to the lowest-energy state, is denoted as
\be
\label{4}
N_0 \; = \; \sup_k n_k \;  .
\ee

The field operator can be expanded over the natural orbitals
\be
\label{5}
 \psi(\br) \; = \; \sum_k a_k \vp_k(\br) \; , \qquad
n_k \; = \; \lgl \; a_k^\dgr \; a_k \; \rgl \;  ,
\ee
separating the ground-state and excited terms,
\be
\label{6}
\psi(\br) \; = \; \psi_0(\br) + \psi_1(\br) \; , \qquad
\psi_0(\br)  \; = \; a_0 \;\vp_0(\br) \; , \qquad \psi_1(\br)  \; = \;
\sum_{k\neq 0} a_k \;\vp_k(\br) .
\ee
Respectively, the number-of-particle operator can be separated into two terms,
$$
\hat N \; = \; \int \psi^\dgr(\br) \; \psi(\br) \; d\br \; = \;
\hat N_0 + \hat N_1 \; ,
$$
\be
\label{7}
\hat N_0  \; = \; a_0^\dgr \; a_0 \; , \qquad
\hat N_1 \; = \; \sum_{k\neq 0} a_k^\dgr \; a_k \;  ,
\ee
thus getting the ground-state occupation number
\be
\label{8}
N_0 \; = \; \lgl \; \hat N_0 \; \rgl \; = \;
\lgl a_0^\dgr \; a_0 \; \rgl \;   .
\ee

Formally separating the ground-state does not necessarily make it representing BEC.
The latter is recovered in the thermodynamic limit, which for a uniform system of the
particle number $N$ in volume $V$ means the limits
\be
\label{9}
N \; \ra \; \infty \; , \qquad  V \; \ra \; \infty \; , \qquad
\frac{N}{V} \; \ra \; const \; .
\ee
The more general form \cite{Yukalov_15} of the thermodynamic limit, valid for arbitrary
nonuniform, as well as for uniform systems, reads as
\be
\label{10}
N \; \ra \; \infty \; , \qquad  A_N \; \ra \; \infty \; , \qquad
\frac{A_N}{N} \; \ra \; const \;   ,
\ee
with $A_N$ being an extensive observable quantity.

The criterion of the BEC existence is the condition
\be
\label{11}
\lim_{N\ra\infty}\; \frac{N_0}{N} \; = \;
\lim_{N\ra\infty}\; \frac{\lgl a_0^\dgr \; a_0\rgl}{N} \; > \; 0 \; ,
\ee
in which the thermodynamic limit is assumed. This criterion is more general than the
formulation of Yang \cite{Yang_16} defining the condensate density as the off-diagonal
limit
\be
\label{12}
 \lim_{r\ra\infty} \rho(\br,0) \; = \; \rho_0 \; > \; 0 \; ,
\ee
since for trapped systems this limit is always zero, independently of existence or
absence of BEC. On the contrary, the criterion (\ref{11}), characterizing the macroscopic 
occupation of a quantum state, is general, being valid for both uniform as well as trapped 
particles \cite{Lieb_2002,Cederbaum_2017}.

\section{Gauge symmetry breaking}

\subsection*{A. Gauge symmetry}

A system is globally gauge symmetric when the number-of-particle operator commutes with
the Hamiltonian, $[\hat{N},H] = 0$. Then the latter is invariant with respect to the
gauge transformation,
\be
\label{13}
\hat U^+_\al \; H \; \hat U_\al \; = \; H \; , \qquad
\hat U_\al \; \equiv \; e^{i\al \hat N} \;  ,
\ee
where $\alpha$ is real-valued. The symmetry is global because $\alpha$ does not depend
on spatial coordinates. In other words, the Hamiltonian is gauge symmetric when it does
not vary under the replacement $\psi \longmapsto \psi e^{i \alpha}$. In what follows,
for short, the word ``global" will be often omitted, keeping in mind that we will always
be talking on global gauge symmetry. 

As an example of a Hamiltonian with global gauge symmetry, we can mention the standard form
of a Hamiltonian for particles with a pair-interaction potential $\Phi(\br-\br')$ in an 
external field $U = U(\br)$, 
$$
\hat H \; = \;
\int \psi^\dgr(\br) \; \left( -\; \frac{\nabla^2}{2m} + U \right) \;
\psi(\br) \; d\br + \frac{1}{2}
\int \psi^\dgr(\br) \; \psi^\dgr(\br') \; \Phi(\br-\br') \;
\psi(\br') \; \psi(\br) \; d\br d\br' \;  .
$$

For a gauge symmetric Hamiltonian, the statistical averages, composed of field-operator 
products, satisfy the equality
\be
\label{14}
\left\lgl \; \prod_{i=1}^m \psi^\dgr(\br_i) \;
\prod_{j=1}^n \psi(\br_j) \; \right\rgl \; = \; 0
\qquad
(m \neq n) \;  ,
\ee
whose proof can be found, e.g., in the review \cite{Yukalov_17}. The averages with
$m = n$ are called ``normal", while those with $m \neq n$, are named ``anomalous".
However, when the gauge symmetry becomes broken, then both nonzero normal and anomalous
averages can survive.

\subsection*{B. Method of quasi-averages}

The procedure of symmetry breaking can be done in several ways. An efficient method of
symmetry breaking is the introduction in the Hamiltonian of small terms breaking the
desired symmetry and removing these terms after the thermodynamic limit
\cite{Bogolubov_5,Bogolubov_6,Bogolubov_7,Kirkwood_18}. For example, assume that a
Hamiltonian $H$ is invariant with respect to some symmetry transformation. One defines
a Hamiltonian
\be
\label{15}
H_\ep\; = \; H + \ep \hat B
\ee
by adding a term in which $\hat{B}$ breaks the symmetry and $\varepsilon$ ia a real-valued
parameter. The statistical average of the operator $A_N$ of an observable is defined as
a {\it quasi-average}
\be
\label{16}
\lim_{\ep\ra 0} \; \lim_{N\ra\infty} \; \frac{\lgl \hat A_N \rgl_\ep}{N} \;   ,
\ee
where the averaging involves the symmetry breaking Hamiltonian (\ref{15}). It is
important to stress that the removal of the symmetry-breaking term, by sending
$\varepsilon$ to zero, has to be done only after the thermodynamic limit. If the order of 
the limits would be reversed, with $\varepsilon$ tending to zero before the thermodynamic 
limit, then the symmetry-breaking term in (15) would disappear, the Hamiltonian symmetry 
would be restored and one would return back to the symmetric Hamiltonian that is not able 
to describe spontaneous symmetry breaking, hence phase transitions.

In the case, where the additional infinitesimal term breaks the gauge symmetry, and
the quasi-average
\be
\label{17}
\lim_{\ep\ra 0} \; \lim_{N\ra\infty} \; \frac{1}{N}
\int \left| \; \lgl \; \psi_0(\br) \; \rgl_\ep \; \right|^2 \; d\br \; > \; 0
\ee
is not zero, one says that there occurs spontaneous gauge symmetry breaking. Condition
(\ref{17}) implies that
\be
\label{18}
\lim_{\ep\ra 0} \; \lim_{N\ra\infty} \;
\frac{|\;\lgl \; a_0 \; \rgl_\ep\; |^2}{N} \; > \; 0 \; .
\ee

Instead of two limiting conditions in the definition of quasi-averages, it is possible
to reduce the consideration to a single thermodynamic limit by introducing the symmetry
breaking Hamiltonian in the form
$$
H_N \; = \; H + \frac{1}{N^\gm} \; \hat B \qquad ( 0 < \gm < 1) \; ,
$$
which allows to define {\it thermodynamic quasi-averages} \cite{Yukalov_19}, for which 
the usual thermodynamic limit is sufficient. Below, we shall use the standard 
quasi-averages (\ref{16}).

\subsection*{C. Bogolubov theorem}

Assume that there is a class of correlation functions $C_\varepsilon[\psi_0,\psi_1]$
composed of the products of the operators $\psi_0$ and $\psi_1$. Bogolubov
\cite{Bogolubov_5,Bogolubov_6} showed that in the thermodynamic limit the operators
$\psi_0$ can be replaced by a non-operator function $\eta$, so that
\be
\label{19}
\lim_{\ep\ra 0} \; \lim_{N\ra\infty} C_\ep[\; \psi_0, \; \psi_1\; ] \; = \;
\lim_{\ep\ra 0} \; \lim_{N\ra\infty} C_\ep[\; \eta, \; \psi_1\; ] \; ,
\ee
with $\eta$ being a minimizer of the free energy
\be
\label{20}
F_\ep \; = \; - T \ln {\rm Tr} \; e^{-\bt H_\ep} \qquad
\left( \bt \equiv \frac{1}{T} \right) \; ,
\ee
hence satisfying the equation
\be
\label{21}
\lim_{\ep\ra 0} \; \lim_{N\ra\infty} \; \frac{\dlt F_\ep}{\dlt\eta}
 \; = \;
\lim_{\ep\ra 0} \; \lim_{N\ra\infty} \; \left\lgl \;
\frac{\dlt H_\ep}{\dlt\eta} \; \right\rgl_\ep \; = \; 0 \;  .
\ee

Thus, one comes to the correlation functions defining the minimizer
\be
\label{22}
\lim_{\ep\ra 0} \; \lim_{N\ra\infty} \;
\lgl \;\psi_0(\br) \; \rgl_\ep \; = \; \eta(\br)
\ee
and the density of particles on the lowest level
\be
\label{23}
\lim_{\ep\ra 0} \; \lim_{N\ra\infty} \;
\lgl \;\psi_0^\dgr(\br) \; \psi_0(\br) \; \rgl_\ep \; = \;
|\; \eta(\br)\; |^2 \;   .
\ee
If the minimizer is not zero, this means that conditions (\ref{17}) and (\ref{18})
are valid, hence the global gauge symmetry is broken. And nonzero $\eta$ defines
the condensate density (\ref{23}).

\subsection*{D. Condensate existence}

The Bogolubov theorem tells us that spontaneous gauge symmetry breaking leads to the
condensate existence, which is characterized by the condensate function $\eta$. For
an equilibrium system, the condensate density is $|\eta|^2 = N_0/N$. The fact that the
global gauge symmetry breaking is a sufficient condition for the condensate existence
can also be shown in a rather simple way, by involving the Cauchy-Schwarz inequality
$$
|\; \lgl \; a_0 \; \rgl_\ep \; |^2 \; \leq \;
\lgl \; a_0^\dgr \; a_0 \; \rgl_\ep \;  .
$$
Using this inequality it is easy to see that, if gauge symmetry is broken, hence condition
(\ref{18}) holds true, then
\be
\label{24}
\lim_{\ep\ra 0} \; \lim_{N\ra\infty} \;
\frac{\lgl\; a_0^\dgr \; a_0 \; \rgl_\ep}{N} \; > \; 0 \; ,
\ee
hence there exists the condensate.

\subsection*{E. Ginibre theorem}

Ginibre \cite{Ginibre_20} proved the following theorem. Consider a symmetry-breaking
grand Hamiltonian $H_\varepsilon = H_\varepsilon[\psi_0,\psi_1]$ and the related grand
potential
\be
\label{25}
\Om_\ep [\; \psi_0 , \psi_1 \; ] \; = \; - T \ln 
{\rm Tr} \exp\{ - \bt H_\ep [ \; \psi_0 , \psi_1 \; ] \} \; .
\ee
Compare this thermodynamic potential with the potential
\be
\label{26}
\Om_\ep [\; \eta , \psi_1 \; ] \; = \; - T \ln 
{\rm Tr} \exp\{ - \bt H_\ep [ \; \eta , \psi_1 \; ] \} \;   ,
\ee
having the same form, except the operator $\psi_0$ being replaced by a non-operator
quantity $\eta$. The latter is defined as a minimizer of the thermodynamic potential,
so that
\be
\label{27}
\frac{\prt}{\prt \eta} \; \Om_\ep [\; \eta , \psi_1 \; ] \; = \;
\left\lgl \; \frac{\prt}{\prt \eta} \; H_\ep [\; \eta , \psi_1 \; ] \;
\right\rgl \; = \; 0 \; .
\ee
Then in the thermodynamic limit the equality is valid:
\be
\label{28}
\lim_{\ep\ra 0} \; \lim_{N\ra\infty} \; \frac{1}{N} \;
\Om_\ep  [\; \psi_0 , \psi_1 \; ] \; = \;
\lim_{\ep\ra 0} \; \lim_{N\ra\infty} \; \frac{1}{N} \; \sup_\eta
\Om_\ep  [\; \eta , \psi_1 \; ] \; .
\ee
Again we see that global gauge symmetry breaking is a sufficient condition for BEC that
is characterized by the condensate density $|\eta|^2 = N_0/N$.

The Ginibre theorem \cite{Ginibre_20} was proven for uniform systems. However Lieb at al. 
\cite{Lieb_2005,Lieb_32} have recently pointed out how this theorem can be straightforwardly 
generalized to nonuniform case.

\subsection*{F. Roepstorff theorem}

A very important theorem has been proved by Roepstorff \cite{Roepstorff_21} (see also 
\cite{Lieb_32}), who showed that the global gauge symmetry breaking necessarily happens 
under BEC. He proved the inequality
\be
\label{29}
\lim_{N\ra\infty} \; \frac{\lgl \; a_0^\dgr a_0 \; \rgl}{N} \; \leq \;
\lim_{\ep\ra 0} \; \lim_{N\ra\infty} \;
\frac{|\;\lgl\; a_0 \;\rgl_\ep \; |^2}{N} \; ,
\ee
which is equivalent to the inequality
\be
\label{30}
 \lim_{N\ra\infty} \; \frac{1}{N} \int
\lgl \; \psi_0^\dgr(\br) \; \psi_0(\br) \; \rgl \; d\br \; \leq \;
\lim_{\ep\ra 0} \; \lim_{N\ra\infty} \;  \frac{1}{N}
\int |\; \lgl \; \psi_0(\br) \; \rgl_\ep \; |^2 \; d\br \;  .
\ee
Note that the left-hand sides of the above inequalities are expressed through
the averages without gauge symmetry breaking. From here it follows that the
appearance of BEC, when condition (\ref{11}) is valid, is necessarily accompanied
by the global gauge symmetry breaking, when condition (\ref{18}) holds true.

\subsection*{G. Bogolubov shift}

The above theorems provide us a rigorous mathematical basis for the understanding
that: {\it The global gauge symmetry breaking is the necessary and sufficient
condition for BEC}. Any approaches or methods, attempting to treat BEC without
gauge symmetry breaking are not correct \cite{Yukalov_73}.

Bogolubov \cite{Bogolubov_5,Bogolubov_6} also proved that the gauge symmetry breaking
can be conveniently realized by the operator shift
\be
\label{31}
\psi(\br) \; = \; \eta(\br) + \psi_1(\br) \; ,
\ee
where $\eta$ is a condensate wave function and $\psi_1$ is the field operator of
non-condensed particles. This method is equivalent to the method of quasi-averages.
The equality (\ref{31}) is an exact canonical transformation, but not an approximation,
as one often states.

The field operator of non-condensed particles satisfies the equality
\be
\label{32}
\lgl \; \psi_1(\br) \; \rgl \; = \; 0
\ee
that has the meaning of a quantum-number conservation condition required for the
conservation of such characteristics as spin or momentum. The condensate function
and the field operator of uncondensed particles are independent variables, because
of which they are to be orthogonal,
\be
\label{33}
 \int \eta^*(\br) \; \psi_1(\br) \; d\br \; = \; 0 \;  .
\ee

The condensate function satisfies the requirement, imposed by the Bogolubov-Ginibre
theorems, according to which this function is normalized to the number of condensed
particles
\be
\label{34}
N_0 \; = \; \int |\; \eta(\br) \; |^2 \; d\br \;   ,
\ee
and this number is prescribed to be a minimizer of thermodynamic potential. From
the other side, the total average number of particles $N$ is also fixed. When $N_0$
and $N$ are fixed, then $N_1=N-N_0$ is also fixed, thus yielding the second
normalization condition
\be
\label{35}
 N_1 \; = \; \lgl \; \hat N_1 \; \rgl \; = \;
\int \lgl \; \psi_1^\dgr(\br) \; \psi_1(\br) \; \rgl \; d\br \;  .
\ee
In that way, for a system with gauge symmetry breaking there are two normalization
conditions fixing either $N_0$, as a thermodynamic potential minimizer, and the total
number of particles $N$, or $N_0$ and $N_1$.

\section{Unitary inequivalent representations}

It is necessary to keep in mind that the field operators for a gauge symmetric system and
for a system with broken gauge symmetry have very different properties, being defined in
different Fock spaces. As soon as gauge symmetry becomes broken, one has to deal with
the related Fock space that is orthogonal to the space associated with a system without
gauge symmetry breaking.

Let the field operator $\psi$ correspond to a gauge symmetric system with a vacuum
$|0 \rangle$, for which
\be
\label{36}
 \psi(\br) \; |\; 0 \; \rgl \; = \; 0 \;  .
\ee
Other states can be created by means of the creation operators $\psi^\dagger$ following
the rule
\be
\label{37}
 |\; \vp \;  \rgl \; = \; \sum_{n=0}^\infty \frac{1}{\sqrt{n!}} \;
\int \vp(\br_1,\br_2,\ldots,\br_n) \; \prod_{i=1}^n
\psi^\dgr(\br_i) \; d\br_i \; | \; 0 \; \rgl \; ,
\ee
where $\varphi$ is a complex function symmetric with respect to its arguments. The closed
linear envelope of these states composes the Fock space
\be
\label{38}
{\cal F}(\psi^\dgr) \; = \; \overline\cL\{ \;  | \; \vp \; \rgl \; \} \;  .
\ee

Accomplishing the Bogolubov shift (\ref{31}), we come to the field operators $\psi_1$,
for which the state $|0\rangle$ is not a vacuum, since $\psi_1 |0\rangle$ is not zero.
Let its vacuum be $|0\rangle_1$, so that
\be
\label{39}
 \psi_1(\br) \; |\; 0 \; \rgl_1 \; = \; 0 \;  .
\ee
Then the closed linear envelope of the states
\be
\label{40}
|\; \vp_1 \;  \rgl \; = \; \sum_{n=0}^\infty \frac{1}{\sqrt{n!}} \;
\int \vp_1(\br_1,\br_2,\ldots,\br_n) \; \prod_{i=1}^n
\psi_1^\dgr(\br_i) \; d\br_i \; | \; 0 \; \rgl_1
\ee
forms the Fock space
\be
\label{41}
{\cal F}(\psi_1^\dgr) \; = \; \overline\cL\{ \;  | \; \vp_1 \; \rgl \; \} \; .
\ee

Noticing that
\be
\label{42}
\psi(\br) \; |\; 0 \; \rgl_1 \; = \; \eta(\br) \; | \; 0 \;\rgl_1 \;   ,
\ee
we see that the vacuum $|0\rangle_1$ is a coherent state with respect to the operators
$\psi$, hence
\be
\label{43}
  | \; 0 \;\rgl_1 \; = \; \exp\left( -\; \frac{1}{2} \; N_0 \right) \;
\exp\left\{ \int \eta(\br) \; \psi^\dgr(\br) \; d\br \;
\right\} \; | \; 0 \; \rgl \;  .
\ee
While from the equality
\be
\label{44}
\psi_1(\br) \; |\; 0 \; \rgl \; = \; - \eta(\br) \; | \; 0 \;\rgl
\ee
it follows that the vacuum $|0\rangle$ is a coherent state with respect to $\psi_1$,
that is
\be
\label{45}
 | \; 0 \; \rgl \; = \;
\exp\left( -\; \frac{1}{2} \; N_0 \right) \;
\exp\left\{ - \int \eta(\br) \; \psi_1^\dgr(\br) \; d\br \;
\right\} \; | \; 0 \; \rgl_1 \; .
\ee
Then it is not difficult to show that
\be
\label{46}
\lgl \; 0 \; | \; 0 \; \rgl_1 \; = \;
\exp\left( -\; \frac{1}{2} \; N_0 \right) \;   ,
\ee
which tends to zero in the thermodynamic limit, when $N_0$ is proportional to
$N \ra \infty$.

In this way, the operators $\psi$ and $\psi_1$ act in different Fock spaces, one being
gauge symmetric, while the other having broken gauge symmetry. These spaces are
orthogonal to each other. Although both types of the field operators possess the same
Bose commutation relations, but, being defined on different Fock spaces, they realize
what is called inequivalent representations of commutation relations
\cite{Umezawa_22,Yukalov_23,Yukalov_24}.

\section{Thermodynamic stability}

A very important point that is often forgotten is the necessity of checking the stability
of the studied system. There is no much reason in accomplishing a lengthy investigation
of the system properties, if this system is unstable and, actually, cannot exist.

The system thermodynamic stability is based on the minimization of a thermodynamic
potential \cite{Kubo_25}, from which the requirements on susceptibilities follow. Specific
heat for Bose systems, as a rule, satisfies the condition of stability, being positive.
For Bose-condensed systems, more important is the behavior of isothermic compressibility
that can be defined by one of the following equivalent expressions, whose convenience
depends on the used statistical ensemble:
$$
\varkappa_T \; = \;
\frac{1}{V} \; \left( \frac{\prt^2 F}{\prt V^2}\right)^{-1}_{TN} \; = \;
-\; \frac{1}{V} \; \left( \frac{\prt^2 G}{\prt P^2}\right)_{TN} \; = \;
-\; \frac{1}{\rho N} \; \left( \frac{\prt^2\Om}{\prt\mu^2}\right)_{TV} \; =
$$
$$
=
-\; \frac{1}{V} \; \left( \frac{\prt V}{\prt P}\right)_{TN} \; = \;
-\; \frac{1}{V} \; \left( \frac{\prt P}{\prt V}\right)^{-1}_{TN} \; =
$$
\be
\label{47}
 = \;
\frac{1}{\rho} \; \left( \frac{\prt\rho}{\prt P}\right)_{TN} \; = \;
\frac{1}{\rho} \; \left( \frac{\prt P}{\prt\rho}\right)^{-1}_{TN} \; = \;
\frac{1}{\rho N} \; \left( \frac{\prt N}{\prt\mu}\right)_{TV} \; = \;
\frac{1}{N} \; \left( \frac{\prt N}{\prt P}\right)_{TV} \;  .
\ee
Here $F=F(T,V,N)$ is free energy, $G=G(T,P,N)$, Gibbs potential, $\Om=\Om(T,V,\mu)$,
grand potential, $P$ is pressure, $\mu$, chemical potential, $T$, temperature, and $\rho$
is density.

The compressibility is connected with the sound velocity by the expression
\be
\label{48}
s^2 \; =\; \frac{1}{m} \; \left( \frac{\prt P}{\prt\rho}\right)_{TN} \; ,
\ee
so that
\be
\label{49}
 \varkappa_T \; = \; \frac{1}{m\rho s^2} \;  .
\ee

From the other side, the compressibility is linked to particle fluctuations,
\be
\label{50}
 \varkappa_T \; = \; \frac{{\rm var}(\hat N)}{\rho TN} \; ,
\ee
where
$$
{\rm var}(\hat N) \; \equiv \; \lgl \; \hat N^2 \; \rgl -
\lgl \; \hat N \; \rgl^2 \; , \qquad
N  \; = \; \lgl \; \hat N \; \rgl \; = \;
\int \lgl \; \psi^\dgr(\br) \; \psi(\br) \; \rgl \; d\br \; .
$$

Thermodynamic stability requires that the isothermic compressibility be non-negative
and finite:
\be
\label{51}
0 \; \leq \; \varkappa_T \; < \; \infty \; .
\ee
Both these requirements have straightforward physical explanations. A negative
compressibility, not minimizing a thermodynamic potential, would result in the system
blow, while the infinite positive compressibility would lead to the system collapse.
Because of equality (\ref{50}), the stability condition (\ref{51}) can be rewritten
as the condition on the relative particle variance:
\be
\label{52}
0 \; \leq \; \frac{{\rm var}(\hat N)}{N}\; < \; \infty \;   .
\ee

The above thermodynamic relations are exact and are valid independently on whether gauge
symmetry is broken or not.

\section{Dynamic stability}

One sometimes distinguishes thermodynamic stability and dynamic stability, understanding
under the latter the study of dynamic structure factor $S({\bf k},\omega)$ and the related
structure factor
\be
\label{53}
 S(\bk) \; = \; \int_{-\infty}^\infty S(\bk,\om) \; \frac{d\om}{2\pi} \; ,
\ee
whose properties are described, e.g., in Refs. \cite{Gurevich_26,Hansen_27,Yukalov_28}.
The dynamic structure factor is the Fourier transform of the density-density correlation
function and it is connected with the response function $\chi$ by the relation
\be
\label{54}
S(\bk,\om) \; = \; - \frac{2{\rm Im}\;\chi(\bk,\om)}{\rho(1+e^{-\bt\om})}
\ee
and by the sum rule
\be
\label{55}
 \int_{-\infty}^\infty \frac{1}{\om} \; S(\bk,\om) \; \frac{d\om}{2\pi} \; = \;
- \; \frac{1}{\rho} \; {\rm Re}\; \chi(\bk,0 ) \;  .
\ee

The long-wave structure factor
\be
\label{56}
S(0) \; = \; 1 + \frac{1}{N} \int \rho(\br) \; \rho(\br') \;
[\; g(\br,\br') - 1 \; ] \; d\br d\br'
\ee
is expressed through the pair correlation function
\be
\label{57}
 g(\br,\br') \; = \;
\frac{\rho_2(\br,\br',\br,\br')}{\rho(\br)\; \rho(\br')}
\ee
and the second-order reduced density matrix
\be
\label{58}
\rho_2(\br,\br',\br,\br') \; = \; \lgl \; \psi^\dgr(\br') \; \psi^\dgr(\br) \;
\psi(\br) \; \psi(\br') \; \rgl \; .
\ee
This gives the formula
\be
\label{59}
  S(0) \; = \; \frac{{\rm var}(\hat N)}{N} \; ,
\ee
which results in the relations
\be
\label{60}
\varkappa_T  \; = \; - \; \frac{1}{\rho^2}\;{\rm Re}\;\chi(0,0) \; = \;
\frac{S(0)}{\rho T} \;  .
\ee

In this way, the study of the dynamic structure factor and the related structure factor
lead to the investigation of the particle variance and isothermic compressibility, hence
to the same stability conditions (\ref{51}) and (\ref{52}). Therefore, if one uses the
same approximation, then both ways produce identical conclusions. The typical mistake is
the use of different approximations that can give different results. For example, one
employs thermodynamic relations, accomplishing calculations with broken gauge symmetry,
while, when studying the structure factor, one falls back to the Hartree approximation
without taking account of symmetry breaking. Of course, the conclusions then will be
different, since the use of the gauge-symmetric Hartree approximation is not correct
for a Bose-condensed system.

\section{Absence of ``grand canonical catastrophe"}

A very wide spread delusion is the belief in the existence of the so-called ``grand
canonical catastrophe". Starting from seventies \cite{Fujiwara_29,Ziff_30}), hundreds 
of papers have claimed the occurrence of this ``catastrophe" (see the history in 
\cite{Kruk_2025}). This delusion comes into play in the following way. For a system 
without gauge symmetry breaking, the variance ${\rm var}(\hat{N}_0)$ for the ground-state 
level has the form $N_0(1+N_0)$. One remembers that, under BEC, $N_0$ becomes of order 
$N$. This means that particle fluctuations are catastrophically large, being proportional 
to $N^2$. Therefore the compressibility is of order $N$ and diverges in the thermodynamic 
limit, thus making the system unstable. One even goes so far as to claim that the grand 
canonical ensemble is not suitable for treating BEC. However, all that is, of course, 
a confusion caused by forgetting to break the gauge symmetry, while BEC happens only 
together with gauge symmetry breaking \cite{Ter_31}. Because of the importance of this 
topic, let us explain this point a bit more in detail.

Consider a Bose gas described by the standard Hamiltonian
\be
\label{61}
 H \; = \; \sum_k \left( \frac{k^2}{2m} - \mu \right) \; a_k^\dgr a_k
\ee
in the grand canonical ensemble, with fixed $T$ and $\mu$. Following the prescription,
explained above, let us break the gauge symmetry by adding to the Hamiltonian the
infinitesimal term $\varepsilon {\hat B}$ getting
\be
\label{62}
H_\ep \; = \; H + \ep \hat B \; = \; \sum_{k\neq 0}
\left( \frac{k^2}{2m} - \mu \right) \; a_k^\dgr a_k - \mu \; a_0^\dgr a_0 +
\ep\; \sqrt{V} \; \left( a_0^\dgr \; e^{i\vp} + a_0 \; e^{-i\vp} \right) \;   ,
\ee
where $\varphi$ is a real phase. By the canonical transformation
\be
\label{63}
 a_0 \; = \; b_0 + \ep \; \frac{\sqrt{V}}{\mu} \; e^{i\vp} \;  ,
\ee
the Hamiltonian (\ref{62}) can be reduced to the diagonal form
\be
\label{64}
 H_\ep \; = \; \sum_{k\neq 0}
\left( \frac{k^2}{2m} - \mu \right) \; a_k^\dgr a_k - \mu \; b_0^\dgr b_0 +
\frac{\ep^2}{\mu}\; V \; .
\ee

With this diagonal Hamiltonian, it is straightforward to find the averages for the
operators $b_0$,
\be
\label{65}
 \lgl \; b_0 \; \rgl_\ep \; = \; 0 \; , \qquad
\lgl \; b_0^\dgr b_0 \; \rgl_\ep \; = \; \frac{z}{1-z} \qquad
\left( z \equiv e^{\bt\mu}\right) \; ,
\ee
where $z\neq 1$, and for the operators $a_k$,
$$
\lgl \; a_0 \; \rgl_\ep \; = \;  \frac{\ep}{\mu} \; \sqrt{V} \; e^{i\vp} \; ,
\qquad
\lgl \; a_0^\dgr a_0 \; \rgl_\ep \; = \;  \frac{z}{1-z} +
\frac{\ep^2}{\mu^2} \; V \; ,
$$
\be
\label{66}
\lgl \; a_k^\dgr a_k \; \rgl_\ep \; = \; \frac{1}{e^{\bt\om_k}-1} \qquad
\left( \om_k \equiv \frac{k^2}{2m} - \mu \right) \;   .
\ee

The variance of the condensate number-of-particle operator,
\be
\label{67}
{\rm var}(\hat N_0) \; \equiv \; \lgl \; \hat N_0^2 \; \rgl_\ep -
\lgl \; \hat N_0 \; \rgl^2_\ep \; = \;
T \;\frac{\prt N_0}{\prt\mu} \;   ,
\ee
where
\be
\label{68}
N_0  \; = \;  \lgl \; a_0^\dgr a_0 \; \rgl_\ep \; = \;
\frac{z}{1-z} + \frac{\ep^2}{\mu^2} \; V \; ,
\ee
takes the form
\be
\label{69}
 {\rm var}(\hat N_0) \; = \;
\frac{z}{(1-z)^2} -\; \frac{2T\ep^2}{\rho\mu^3} \; N \; .
\ee
Accomplishing the thermodynamic limit gives
\be
\label{70}
 \lim_{N\ra\infty} \; \frac{{\rm var}(\hat N_0) }{N} \; = \;
- \; \frac{2T}{\rho\mu^3} \; \ep^2 \; .
\ee
And sending $\varepsilon$ to zero reduces the relative variance to zero,
\be
\label{71}
\lim_{\ep\ra 0} \; \lim_{N\ra\infty}
\frac{{\rm var}(\hat N_0) }{N} \; = \;  0 \;    .
\ee

Thus, when correctly treating the problem, no ``grand canonical catastrophe" exists.
The condensate fluctuations are not only normal, without any catastrophic behavior,
but even are absent at al.

If there is no gauge symmetry breaking, then there is no BEC, hence $N_0$ cannot be
proportional to $N$, and no problem arises, while when BEC occurs, hence the gauge
symmetry becomes broken, then there are no condensate fluctuations at al. The
``catastrophe" appears only due to incorrect calculations, by treating BEC without
global gauge symmetry breaking.

The situation becomes absolutely transparent and even trivial, if one remembers that
breaking gauge symmetry, with using quasi-averages, is equivalent to the use of the
Bogolubov shift (\ref{31}), where the condensate number-of-particle operator is just
${\hat N}_0 \equiv N_0 {\hat 1}$, which evidently gives zero variance, 
${\rm var}(\hat N_0) = {\rm var}(N_0 {\hat 1}) = 0$. More details can be found in 
\cite{Yukalov_32}.

\section{Instability of ideal Bose gas}

One often compares the behavior of interacting Bose gas with ideal Bose gas. Some of
observable quantities for an ideal gas can be close to those for a weakly interacting
gas. However, this does not mean that all quantities, without exception, have to be
close to those corresponding to the interacting Bose gas. It is necessary to keep in
mind that the ideal Bose gas and interacting Bose gas pertain to different universality
classes, whose properties in some aspects can be drastically different. The ideal Bose
gas pertains to the Gaussian class of models, while the interacting Bose system,
independently of the strength of finite interactions, pertains to the class named $XY$
or $O(2)$. In particular, it turns out that the properties of their particle fluctuations,
hence conditions of stability, are rather different.

In a Bose-condensed system, the condensate does not exhibit particle fluctuations, as is
shown in the previous section. This implies that all particle fluctuations are due to
non-condensed particles,
\be
\label{72}
{\rm var}(\hat N) \; = \; {\rm var}(\hat N_1) \;  .
\ee
For the latter, we have
\be
\label{73}
{\rm var}(\hat N_1) \; = \; T \; \frac{\prt N_1}{\prt\mu} \; = \;
\frac{V}{\lbd_T^3} \; g_{1/2}(1) \; ,
\ee
where
\be
\label{74}
 N_1 \; \equiv \; \frac{V}{\lbd_T^3}\; g_{3/2}(z) \qquad
\left( \lbd_T \equiv \sqrt{\frac{2\pi}{m T} } \right) \;  .
\ee
Here $g_n(z)$ is the Bose integral that converges not for all $n$. In order to
understand the character of its behavior it is necessary to take into account
that, for a finite system of volume $V$, there exists the minimal energy
$\ep_{min}=k^2_{min}/2m$ corresponding to the minimal wave vector $k_{min}=2\pi/L$,
with $L = V^{1/3}$. The existence of the minimal energy requires that the integral
$g_n(z)$ be limited from below by the dimensionless quantity $u_0 = \ep _{min}/T$.
This suggests to introduce \cite{Yukalov_15} the generalized Bose integral
\be
\label{75}
g_n(z) \; = \; \frac{1}{\Gm(n)} \int_{u_0}^\infty
\frac{z u^{n-1}}{e^u-z} \; du \qquad
\left( u_0 = \pi\; \frac{\lbd_T^2}{L^2}\right) \;   .
\ee

For the variance of non-condensed particles, we get
\be
\label{76}
{\rm var}(\hat N_1) \; = \; \frac{2}{\pi\lbd_T^4}\; V^{4/3} \; ,
\ee
so that the isothermic compressibility is divergent,
\be
\label{77}
 \varkappa_T \; = \; \frac{m^2 T}{2\pi^3\rho^2}\;V^{1/3} \; ,
\ee
hence the system is unstable. Exactly the same result follows from the quantum
calculations of Ziff \cite{Ziff_30}.

The instability of the ideal uniform Bose gas in three dimensions is not so much
surprising, since this is, actually, an imaginary object that, strictly speaking,
does not exist. In a real gas, there are always some interactions. And interactions
of any strength, even being infinitesimally small, but finite, make the gas stable.
The infinite compressibility implies that the gas will collapse, and in the process
of the collapse, sooner or later, there will necessarily appear collisions and
interactions, the more violent, the more the system is compressed. So that interactions 
will have to arise under realistic physical conditions.

\section{Role of dimensionality}

Stability of the ideal uniform Bose gas depends on the space dimensionality. Particle
fluctuations of condensate do not exist in any dimensions, so that all particle
fluctuations are caused by the fluctuations of non-condensed particles, whose number is
\be
\label{78}
 N_1 \; = \; \frac{V}{\lbd_T^d} \; g_{d/2}(z) \;  .
\ee

The condensation occurs below the critical temperature
\be
\label{79}
T_c \; = \;
\frac{2\pi}{m} \; \left[ \; \frac{\rho}{g_{d/2}(1)} \; \right]^{2/d} \; .
\ee
For large $N$, the temperature $T_c$ scales as
$$
T_c \; \propto \; \frac{1}{N} \qquad ( d = 1) \; ,
$$
$$
T_c \; \propto \; \frac{1}{\ln N} \qquad ( d = 2) \; ,
$$
\be
\label{80}
T_c \; \propto \;const \qquad ( d = 3) \;  .
\ee

Above $T_c$, particle fluctuations, depending on the spatial dimensionality, are
described by the reduced variance
$$
\frac{{\rm var}(\hat N)}{N} \; = \; - \; \frac{2z}{(1-z)^2 N} \qquad ( d = 1) \; ,
$$
$$
\frac{{\rm var}(\hat N)}{N} \; = \; - \; \frac{\pi z}{(1-z)^2 N} \qquad ( d = 2) \; ,
$$
\be
\label{81}
\frac{{\rm var}(\hat N)}{N} \; = \;  \frac{1}{\rho\lbd_T^3} \; g_{1/2}(z)
\qquad ( d = 3) \; .
\ee
The compressibility for $d=1$ and $d=2$ is negative, which tells us that the system is
unstable. Above $T_c$, the ideal gas is stable only for $d > 2$.

Below $T_c$, there appears Bose condensate with the temperature dependence of the
condensate fraction
\be
\label{82}
 n_0 \; = \; 1 - \left( \frac{T}{T_c} \right)^{d/2} \qquad
( T \leq T_c ) \;  .
\ee
The relative particle variance, depending on dimensionality, is given by the expressions
$$
\frac{{\rm var}(\hat N)}{N} \; = \; \frac{2L^3}{3\pi^2 \rho \lbd_T^4}
\qquad ( d = 1) \; ,
$$
$$
\frac{{\rm var}(\hat N)}{N} \; = \; \frac{L^2}{\pi \rho\lbd_T^4}
\qquad ( d = 2) \; ,
$$
$$
\frac{{\rm var}(\hat N)}{N} \; = \; \frac{4L}{\pi \rho\lbd_T^4}
\qquad ( d = 3) \; ,
$$
$$
\frac{{\rm var}(\hat N)}{N} \; = \;
\frac{1}{\rho\lbd_T^4} \; \ln \left( \pi \; \frac{\lbd_T^2}{L^2} \right)
\qquad ( d = 4) \; ,
$$
\be
\label{83}
 \frac{{\rm var}(\hat N)}{N} \; = \;
\frac{1}{\rho\lbd_T^5} \; g_{3/2}(1) \qquad ( d = 5) \; .
\ee
The particle-number scaling is
$$
\frac{{\rm var}(\hat N)}{N} \; \propto \; N^3
\qquad ( d = 1) \; ,
$$
$$
\frac{{\rm var}(\hat N)}{N} \; \propto \; N
\qquad ( d = 2) \; ,
$$
$$
\frac{{\rm var}(\hat N)}{N} \; \propto \; N^{1/3}
\qquad ( d = 3) \; ,
$$
$$
\frac{{\rm var}(\hat N)}{N} \; \propto \; \ln N
\qquad ( d = 4) \; ,
$$
\be
\label{84}
\frac{{\rm var}(\hat N)}{N} \; \propto \; const
\qquad ( d = 5) \;   .
\ee
This shows that the Bose gas below $T_c$ is stable for $d > 4$. That is, the critical
dimension for the stability  of normal Bose gas is twice smaller than the critical
dimension for the stability of condensed Bose gas.

\section{Power-law trapping potentials}

For trapped gases, stability also depends on the trap shape. The most often considered
are power-law trapping potentials of the form
\be
\label{85}
U(\br) \; = \;
\sum_{\al=1}^d \frac{\om_\al}{2} \; \left| \; \frac{r_\al}{l_\al} \; \right|^{n_\al}
\qquad
\left( l_\al \equiv \frac{1}{\sqrt{m\om_\al} } \right) \;   ,
\ee
with $d$ being space dimensionality. The effective trap frequency and trap length are
\be
\label{86}
 \om_0 \; \equiv \; \left( \prod_{\al=1}^d \om_\al \right)^{1/d} \; ,
\qquad
l_0 \; \equiv \; \left( \prod_{\al=1}^d \om_\al \right)^{1/d} \; .
\ee

An important quantity characterizing power-law potentials is the confining dimension
\cite{Yukalov_15}
\be
\label{87}
  D \; \equiv \; \frac{d}{2} + \sum_{\al=1}^d \frac{1}{n_\al} \; .
\ee
The finiteness of the trap is taken into account by setting the minimal energy
$\varepsilon_{min} = \omega_0/2$. The generalized Bose function \cite{Yukalov_15}
\be
\label{88}
  g_n(z) \; = \;
\frac{1}{\Gm(n)} \int_{u_0}^\infty \frac{z u^{n-1}}{e^u -z} \; du
\qquad \left( u_0 \equiv \frac{\om_0}{2T} \right)
\ee
defines the main thermodynamic quantities, for instance the number of non-condensed
particles
\be
\label{89}
 N_1 \; = \; \frac{T^D}{\gm_D} \; g_D(z) \;   .
\ee

Explicit calculations of the quantities, typical of the power-law potentials, are
a bit cumbersome \cite{Yukalov_32}, and here we present only some final results
related to the system stability.

The critical temperature is
\be
\label{90}
T_c \; = \; \left[ \; \frac{\gm_D N}{g_D(1)} \; \right]^{1/D} \;   ,
\ee
where
\be
\label{91}
 \gm_D \; \equiv \; \frac{\pi^{d/2}}{2^D}
\prod_{\al=1}^d \frac{\om_\al^{1/2+1/n_\al}}{\Gm(1+1/n_\al)} \;  .
\ee
For different confining dimensions, the critical temperature becomes
$$
T_c \; = \;
\frac{\sqrt{\pi}\; (1 -D)\Gm(D)}{\Gm(1+1/n)} \; \left( \frac{\om_0}{2}\right) \; N
\qquad
(D < 1 , \; d = 1) \; ,
$$
$$
T_c \; = \;
\frac{N\om_0}{\ln(2N)} \qquad (D = 1 , \; d = 1, \; n=2) \; ,
$$
\be
\label{92}
T_c \; = \;
\left[\; \frac{\gm_D N}{\zeta(D)} \; \right]^{1/D} \qquad (D > 1 ) \; .
\ee

The effective thermodynamic limit has to be understood in the sense of the limiting
behavior (\ref{10}), in which $A_N$ is an extensive quantity. Taking for $A_N$ the
system energy
\be
\label{93}
E_N \; = \; \frac{D}{\gm_D} \; g_{D+1}(z) \; T^{D+1} \; ,
\ee
and considering the limit
\be
\label{94}
 N \; \ra \; \infty \; , \qquad E_N \; \ra \; \infty \; , \qquad
\frac{E_N}{N} \; \ra \; const \;  ,
\ee
leads to the thermodynamic limit
\be
\label{95}
N \; \ra \; \infty \; , \qquad \gm_D \; \ra \;0 \; , \qquad
N\gm_D \; \ra \; const \;   .
\ee
For the symmetric potentials, for which $n_\alpha = n$, the thermodynamic limit becomes
\be
\label{96}
N \; \ra \; \infty \; , \qquad \om_0 \; \ra \; 0 \; , \qquad
N \om_0^D \; \ra \; const \;   .
\ee

For different confining dimensions, the critical temperature scales with respect to the
number of particles as
$$
T_c \; \propto \; \frac{1}{N^{(1/D)-1}} \qquad ( D < 1) \; ,
$$
$$
T_c \; \propto \; \frac{1}{\ln N} \qquad ( D = 1) \; ,
$$
\be
\label{97}
 T_c \; \propto \; const \qquad ( D > 1 ) \; .
\ee

Above the critical temperature $T_c$ the reduced variance, depending on the confining
dimension, reads as
$$
\frac{{\rm var}(\hat N)}{N} \; = \; - \;
\frac{z T^D}{(1-z)^2\Gm(1+D) N\gm_D} \;
\left( \frac{\om_0}{2T}\right)^D \qquad ( D < 1) \; ,
$$
$$
\frac{{\rm var}(\hat N)}{N} \; = \; - \;
\frac{z T}{(1-z)^2 N\gm_D} \;
\left( \frac{\om_0}{2T}\right) \qquad ( D = 1) \; ,
$$
\be
\label{98}
\frac{{\rm var}(\hat N)}{N} \; = \;
\frac{T^D}{ N\gm_D} \; g_{D-1}(z) \qquad ( D > 1) \;   .
\ee
This shows that for $T > T_c$ the gas stability happens for $D > 1$.

Below the critical temperature, the condensate fraction formally exists for all confining
dimensions,
\be
\label{99}
n_0 \; = \; 1 - \left( \frac{T}{T_c} \right)^D \qquad ( T \leq T_c) \;   ,
\ee
although not always this has sense because of the system instability. The system stability
is characterized by the reduced variance
$$
\frac{{\rm var}(\hat N)}{N} \; = \; \frac{T^D}{N\gm_D} \; \left[\;
\frac{1}{(2-D)\Gm(D-1)} + \frac{1}{\Gm(D) } \; \right] \;
\left( \frac{2T}{\om_0}\right)^{2-D} \; \qquad ( D < 2) \; ,
$$
$$
\frac{{\rm var}(\hat N)}{N} \; = \; \frac{T^2}{N\gm_2} \;
\ln \left( \frac{2T}{\om_0}\right) \; \qquad ( D = 2) \; ,
$$
\be
\label{100}
\frac{{\rm var}(\hat N)}{N} \; = \;
\frac{T^D}{N\gm_D} \; \zeta(D-1) \; \qquad ( D > 2) \;   .
\ee
The particle-number scaling
$$
\frac{{\rm var}(\hat N)}{N} \; \propto \; N^{(2-D)/D} \qquad ( D < 2) \; ,
$$
$$
\frac{{\rm var}(\hat N)}{N} \; \propto \; \ln N \qquad ( D = 2) \; ,
$$
\be
\label{101}
\frac{{\rm var}(\hat N)}{N} \; \propto \; const \qquad ( D > 2) \; ,
\ee
shows that the ideal Bose-condensed gas in a power-law potential at $T < T_c$ is stable
for $D > 2$. That is, the critical confining dimension for the stability of the condensed
Bose gas (at $T < T_c$) is twice larger than the critical dimension for the stability of
the normal gas (at $T > T_c$).

\section{Resolution of Hohenberg-Martin dilemma}

In the theory of Bose-condensed systems there is a long-standing problem emphasized
by Hohenberg and Martin \cite{Hohenberg_33}, who named it ``conserving versus gapless
approaches". The problem is as follows. Below the critical temperature of BEC, where the
global gauge symmetry is broken, the theoretical description becomes not self-consistent,
suffering either from the appearance of an unphysical gap in the spectrum of collective
excitations, or from the violation of thermodynamic relations. As a consequence of the
infringement of thermodynamic relations, there arise several unphysical conclusions, such
as the BEC phase transition becoming of first order, instead of the second order, and
incorrect behavior of thermodynamic quantities contradicting each other. A detailed
discussion of the Hohenberg-Martin dilemma is given in Refs. \cite{Andersen_34,Yukalov_35}.

The dilemma appears because not all conditions, uniquely characterizing the system with
broken gauge symmetry, have been taken into account. The resolution of this dilemma can
be done in the frame of a representative statistical ensemble taking account of all
conditions imposed on the Bose system \cite{Yukalov_35,Yukalov_36}. The idea of the
representative ensembles, also called generalized Gibbs ensembles, comes back to
Gibbs \cite{Gibbs_37,Gibbs_38}, Ter Haar \cite{Ter_39}, and Jaynes
\cite{Jaynes_40,Jaynes_41}. For a Bose system with broken gauge symmetry the
representative ensemble has been introduced in Refs. \cite{Yukalov_35,Yukalov_36}.

Below the critical temperature $T_c$, where BEC happens, the global gauge symmetry
becomes broken. The symmetry breaking, as explained above, can be broken, e.g., by 
resorting to the method of quasi-averages, introducing infinitesimal sources breaking the
symmetry, or by other equivalent methods. The most convenient way of the gauge symmetry
breaking is the use of the Bogolubov shift (\ref{31}). Then, as is explained in Sec. IV,
all operations are to be accomplished in the Fock space $\mathcal{F}(\psi^\dagger_1)$.
The condensate wave function $\eta$ and the operator of non-condensed particles $\psi_1$
satisfy the quantum number-conservation condition
\be
\label{102}
\lgl \; \psi_1(\br) \; \rgl \; = \; 0 \; , \qquad
\eta(\br) \; = \; \lgl \; \psi(\br) \; \rgl \;   .
\ee
The number of condensed particles (\ref{34}) is prescribed by the Bogolubov-Ginibre
condition of thermodynamic-potential minimization
\be
\label{103}
 \frac{\dlt\Om}{\dlt\eta} \; = \;
\left\lgl \;  \frac{\dlt H}{\dlt\eta} \; \right\rgl \; = \; 0 \; ,
\ee
which for a uniform equilibrium system is equivalent to the condition
\be
\label{104}
\frac{\prt\Om}{\prt N_0} \; = \;
\left\lgl \;  \frac{\prt H}{\prt N_0} \; \right\rgl \; = \; 0 \;   .
\ee

As far as the total average number of particles $N$ is assumed to be fixed, then there
are two conditions imposed on the number of particles, the total number $N$ and the
Bogolubov-Ginibre condition requiring that $N_0$ be a minimizer of a thermodynamic
potential. Since $N = N_0 + N_1$, there is no difference which of the two numbers one
prefers to fix, either $N_0$ and $N$, or $N_1$ and $N$, or $N_0$ and $N_1$. Respectively,
it is possible to define two chemical potentials, say $\mu_0$ and $\mu_1$, such that
\be
\label{105}
\mu_0 \; = \; \frac{\prt F}{\prt N_0} \; , \qquad
\mu_1 \; = \; \frac{\prt F}{\prt N_1} \;   ,
\ee
where $F$ is the system free energy.

It is necessary to stress that no other restrictions are imposed on the values of the
chemical potentials. Thus, from the extremization of the free energy, one has
\be
\label{106}
\dlt F \; = \; \frac{\prt F}{\prt N_0} \; \dlt N_0 +
\frac{\prt F}{\prt N_1} \; \dlt N_1 \; = \; 0 \;   .
\ee
In view of (\ref{105}), this gives
\be
\label{107}
 \mu_0 \; \dlt N_0 + \mu_1 \; \dlt N_1 \; = \; 0 \;  .
\ee
And the relation $N_1 = N - N_0$ yields
\be
\label{108}
(\mu_0 - \mu_1) \; \dlt N_0 \; = \; 0 \; .
\ee
But $N_0$ is prescribed by the Bogolubov-Ginibre minimization condition, hence
$\delta N_0 = 0$, so that condition (\ref{108}) imposes no constraint on the values of
the chemical potentials $\mu_0$ and $\mu_1$, for instance, not requiring their equality.

The chemical potential $\mu$, associated with the total number of particles $N$, is given
by the condition
\be
\label{109}
 \mu \; = \; \frac{\prt F}{\prt N} \; = \;
\frac{\prt F}{\prt N_0} \; \frac{\prt N_0}{\prt N}
+  \frac{\prt F}{\prt N_1} \; \frac{\prt N_1}{\prt N} \; ,
\ee
which, taking account of the definition for the particle fractions
\be
\label{110}
 n_0 \; \equiv \; \frac{N_0}{N} \; , \qquad
  n_1 \; \equiv \; \frac{N_1}{N} \; ,
\ee
results in
\be
\label{111}
\mu \; = \; \mu_0 n_0 + \mu_1 n_1 \;   .
\ee

The number-conservation condition (\ref{102}) can be represented in the standard
form of the average
\be
\label{112}
\lgl \; \hat\Lbd \; \rgl \; = \; 0
\ee
of a Hermitian operator
\be
\label{113}
\hat\Lbd \; = \; \int \left[\; \lbd(\br) \; \psi_1^\dgr(\br) +
\lbd^*(\br) \; \psi_1(\br) \; \right] \; d\br \;  .
\ee
To preserve condition (\ref{112}), the Lagrange multipliers $\lambda$ have to cancel
in the Hamiltonian the terms linear in $\psi_1$.

For an equilibrium system, the statistical operator ${\hat \rho}$ can be obtained by
the standard procedure of conditional minimization of the information functional
$$
I[\; \hat\rho \; ] \; = \; {\rm Tr} \; \hat\rho \; \ln\hat\rho +
\lbd_0 \; ({\rm Tr}\; \hat\rho -1 )  + \bt \; ({\rm Tr}\; \hat\rho \hat H - E ) +
$$
\be
\label{114}
+ \bt \mu_0 \; (N_0 - {\rm Tr}\; \hat\rho \hat N_0 ) +
\bt \mu_1 \; (N_1 - {\rm Tr}\; \hat\rho \hat N_1 ) -
\bt \; {\rm Tr}\; \hat\rho \hat\Lbd \: ,
\ee
taking into consideration all conditions imposed on the system, the normalization
of the statistical operator ${\hat \rho}$, definition of an energy operator ${\hat H}$,
the given numbers of particles $N_0=\lgl{\hat N}_0\rgl$ and $N_1=\lgl{\hat N}_1\rgl$,
where ${\hat N}_0=N_0{\hat 1}$, and the quantum number-conservation condition (\ref{112}).
Minimizing the information functional (\ref{114}) yields the statistical operator
\be
\label{115}
\hat\rho \; = \; \frac{1}{Z} \; e^{-\bt H} \qquad
\left( Z \equiv {\rm Tr}\; e^{-\bt H} \right) \;   ,
\ee
with the grand Hamiltonian
\be
\label{116}
H \; = \; \hat H - \mu_0 N_0 - \mu_1 \hat N_1 - \hat\Lbd \;   .
\ee

The overall approach is straightforwardly extended \cite{Yukalov_24,Yukalov_42,Yukalov_43}
to nonequilibrium Bose-condensed systems. The equation for the condensate wave function
takes the form
\be
\label{117}
  i \; \frac{\prt}{\prt t} \; \eta(\br,t) \; = \;
\left\lgl \; \frac{\dlt H}{\dlt\eta^*(\br,t)} \; \right\rgl \; ,
\ee
and the equation for the field operator of non-condensed particles reads as
\be
\label{118}
  i \; \frac{\prt}{\prt t} \; \psi_1(\br,t) \; = \;
 \frac{\dlt H}{\dlt\psi_1^\dgr(\br,t)}  \;  .
\ee
The latter is identical to the Heisenberg equation of motion due to the equality
\cite{Yukalov_42,Yukalov_43}
\be
\label{119}
 \frac{\dlt H}{\dlt\psi_1^\dgr(\br,t)} \; = \; [\; \psi_1(\br,t) , \; H \; ] \;  .
\ee

\section{Condensate wave function}

The condensate wave function satisfies equation (\ref{117}), in which $H$ is the grand
Hamiltonian (\ref{116}). To get the explicit form of the equation, one needs to specify
the energy Hamiltonian $\hat{H}$, for which one accepts the standard form
\be
\label{120}
 \hat H \; = \;
\int \psi^\dgr(\br) \; \left( -\; \frac{\nabla^2}{2m} + U \right) \;
\psi(\br) \; d\br + \frac{1}{2}
\int \psi^\dgr(\br) \; \psi^\dgr(\br') \; \Phi(\br-\br') \;
\psi(\br') \; \psi(\br) \; d\br d\br' \;  .
\ee
Here $U = U({\bf r},t)$ is an external potential and $\Phi({\bf r})$ is an interaction
potential. To correctly describe a Bose-condensed system, the global gauge symmetry has
to be broken. Employing the Bogolubov shift $\psi = \eta + \psi_1$, we come to the Fock
space $\mathcal{F}(\psi^\dagger_1)$ with broken gauge symmetry. To write the
condensate-function equation in a compact form, let us introduce several notations.

Let us define the first-order density matrix
\be
\label{121}
\rho_1(\br,\br') \; = \; \lgl \; \psi_1^\dgr(\br') \; \psi_1(\br) \; \rgl \; ,
\ee
which is usually termed the normal average, as far as there exists an anomalous average
\be
\label{122}
\sgm_1(\br,\br') \; = \; \lgl \; \psi_1(\br') \; \psi_1(\br) \; \rgl \; .
\ee
The diagonal elements of (\ref{121}) and (\ref{122}) are
\be
\label{123}
\rho_1(\br) \; = \; \rho_1(\br,\br) \; , \qquad
\sgm_1(\br) \; = \; \sgm_1(\br,\br) \;   .
\ee
The total particle density is
\be
\label{124}
 \rho(\br) \; = \; \rho_0(\br) + \rho_1(\br) \; , \qquad
\rho_0(\br) \; = \; |\; \eta(\br) \; |^2 \;  .
\ee
There also happens the triple anomalous average
\be
\label{125}
 \xi_1(\br,\br') \; \equiv \;
\lgl \; \psi_1^\dgr(\br') \; \psi_1(\br') \; \psi_1(\br) \; \rgl \; .
\ee

With the above notations, the condensate-function equation (\ref{117}) reads as
$$
i \; \frac{\prt}{\prt t} \; \eta(\br,t) \; = \;
\left( - \; \frac{\nabla^2}{2m} + U - \mu_0 \right) \; \eta(\br,t) +
$$
\be
\label{126}
 +
\int \Phi(\br-\br') \; \left[ \; \rho(\br',t) \; \eta(\br,t) +
 \rho_1(\br,\br',t) \; \eta(\br',t) +
\sgm_1(\br,\br',t) \; \eta^*(\br',t) + \xi(\br,\br',t) \; \right]\; d\br' \; .
\ee
Here, to stress that the equation is valid for any nonequilibrium system, the
dependence on time is restored. This is the general form of the condensate-function
equation, containing no approximations. For an equilibrium system, the dependence
on time vanishes.

If we consider the ultimate case, where temperature is zero, particle interactions
are asymptotically weak, there are no external forces, and even there are no quantum
fluctuations, so that all particles are Bose-condensed, while there are no non-condensed
particles, which implies that $\psi_1$ is set to zero, then Eq. (\ref{126}) reduces to
the equation
\be
\label{127}
i\;\frac{\prt}{\prt t} \; \eta(\br,t) \; = \;
\left[\; - \; \frac{\nabla^2}{2m} + U - \mu_0 + \int \Phi(\br-\br') \;
\left| \; \eta(\br',t) \;\right|^2 \; d\br' \; \right]\; \eta(\br,t) \;   .
\ee
The meaning of this equation becomes evident, if we average Eq. (\ref{117}) over the
vacuum state, so that
\be
\label{128}
i\;\frac{\prt}{\prt t} \; \eta(\br,t) \; = \;
{}_1\lgl \; 0 \; | \; \frac{\dlt H}{\dlt \eta^*(\br,t)} \; | \; 0 \;
\rgl_1 \;   .
\ee
Then, since $(\eta + \psi_1)|0\rangle_1 = \eta |0\rangle_1$, we get exactly the same
equation (\ref{127}). Therefore Eq. (\ref{127}) is the equation for the vacuum field.

This vacuum-field equation was advanced by Bogolubov \cite{Bogolubov_44} in 1949.
Since then, it has been republished numerous times
(e.g. \cite{Bogolubov_5,Bogolubov_6,Bogolubov_7}). Bogolubov called it the condensate
wave function equation in the semi-classical approximation.

From the mathematical point of view, this is a nonlinear Schr\"{o}dinger equation
\cite{Malomed_45}. It has been studied by many authors starting from the pioneering
suggestion by Bogolubov \cite{Bogolubov_44}, detailed investigation by Gross
\cite{Gross_46,Gross_47,Gross_48,Gross_49,Gross_50}, and the papers by
Pitaevskii \cite{Pitaevskii_51}, and Wu \cite{Wu_52}. Gross, similarly to Bogolubov,
considered this equation as a semi-classical approximation for the condensate wave
function.

The same form of the condensate equation (\ref{127}) can be derived in the Schr\"{o}dinger
representation by assuming that the many-particle wave function is a product of
single-particle wave functions, all being in the same ground state \cite{Pethick_53}.
A generalization of the product-type wave function is a Hartree approximation for the
wave function, defined as a sum of products of single-particle wave functions in different 
states \cite{Alon_54}. Such representations, describing a finite number of particles, serve
as models of quasi-condensate, while a real condensate requires the thermodynamic limit.

\section{Self-consistent mean-field approach}

A very widespread misconception is to assume that equation (\ref{127}) is a mean-field
approximation and what is beyond this equation is a beyond-mean-field approximation.
As is explained in the previous section, the vacuum-field equation (\ref{127}) describes
a marginal case of a Bose system under zero temperature, asymptotically weak particle
interactions, absence of any external forces, absence of quantum fluctuations, all
particles being Bose-condensed, and without any non-condensed particles. This is a
vacuum-field equation.

The genuine mean-field approximation is the Hartree-Fock-Bogolubov (HFB) approximation.
It is well known \cite{Andersen_34,Yukalov_35} that the standard HFB approximation is
not self-consistent, exhibiting unphysical gap in the excitation spectrum. But employing
a representative ensemble of Sec. XI resolves the Hohenberg-Martin dilemma and makes
the HFB approach self-consistent. Let us illustrate some results of the self-consistent
approach for a uniform system \cite{Yukalov_24,Yukalov_35,Yukalov_43,Yukalov_62}.

For a uniform system, the field operators $\psi_1$ can be expanded over plane waves,
\be
\label{129}
 \psi_1(\br) \; = \; \frac{1}{\sqrt{V}} \sum_{k\neq 0} a_k e^{i\bk\cdot\br} \;  .
\ee
The interaction potential is assumed to be Fourier transformable,
\be
\label{130}
\Phi(\br) \; = \; \frac{1}{V} \sum_k \Phi_k \; e^{i\bk\cdot\br} \; ,
\qquad
\Phi_k \; = \;  \int \Phi(\br) \; e^{-i\bk\cdot\br} \; d\br \; .
\ee
The condensate function becomes $\eta = \sqrt{\rho_0}$ and all densities,
$\rho_0$, $\rho_1$, and $\rho$ are constant with respect to the spatial variables.
The normal and symmetry-broken averages, respectively, are
\be
\label{131}
\rho_1(\br,\br') \; = \;
\frac{1}{V} \sum_{k\neq 0} n_k \; e^{i\bk\cdot(\br-\br')} \; ,
\qquad
\sgm_1(\br,\br') \; = \;
\frac{1}{V} \sum_{k\neq 0} \sgm_k \; e^{i\bk\cdot(\br-\br')} \;   ,
\ee
where
$$
n_k \; = \; \lgl \; a_k^\dgr\; a_k \; \rgl \; = \; \frac{\om_k}{2\ep_k} \;
\coth\left( \frac{\ep_k}{2T} \right) - \; \frac{1}{2} \; ,
$$
\be
\label{132}
\sgm_k \; = \; \lgl \; a_k\; a_{-k} \; \rgl \; = \; -\; 
\frac{\Dlt_k}{2\ep_k} \;
\coth\left( \frac{\ep_k}{2T} \right) \; .
\ee
Here the notations are used:
\be
\label{133}
\om_k \; = \; \frac{k^2}{2m} + \Dlt + \rho_0 \; (\Phi_k - \Phi_0 ) +
\frac{1}{V} \sum_{p\neq 0} n_p \;( \Phi_{k+p}- \Phi_p )
\ee
and
$$
\Dlt_k \; = \;
\rho_0 \; \Phi_k + \frac{1}{V} \sum_{p\neq 0} \sgm_p \; \Phi_{k+p} \; ,
$$
\be
\label{134}
\Dlt \; = \; \lim_{k\ra 0} \Dlt_k \; = \; \rho_0 \; \Phi_0 +
\frac{1}{V} \sum_{p\neq 0} \sgm_p \; \Phi_p \;   .
\ee

The spectrum of excitations
\be
\label{135}
\ep_k \; = \; \sqrt{\om_k^2 - \Dlt_k^2}
\ee
is gapless being at long waves acoustic,
\be
\label{136}
\ep_k \; \simeq \; ck \qquad ( k \ra 0 ) \;   ,
\ee
having the sound velocity
\be
\label{137}
c \; = \; \sqrt{\frac{\Dlt}{m^*} } \;   ,
\ee
with the effective mass
$$
m^* \; = \; \frac{m}{1+\frac{m}{V} \sum_p (n_p-\sgm_p)\Phi_p''} \;   ,
$$
and $\Phi''_p$ is the second derivative with respect to $p$.

Chemical potentials
\be
\label{138}
\mu_0 \; = \;
\rho \; \Phi_0 + \frac{1}{V} \sum_{k\neq 0} ( n_k + \sgm_k) \; \Phi_k \; ,
\qquad
\mu_1 \; = \;
\rho \; \Phi_0 + \frac{1}{V} \sum_{k\neq 0} ( n_k - \sgm_k) \; \Phi_k \;
\ee
guarantee the minimization of the thermodynamic potential and the Hugenholtz-Pines
relation \cite{Hugenholtz_55}. As is evident, the two chemical potentials are
different. As is elucidated in Sec. XI, they do no need to coincide.

It is worth emphasizing that practically all works on Bose-condensed systems employ either 
semi-classical approximation in the form of the nonlinear Schr\"{o}dinger equation, at zero 
temperature, or a kind of Bogolubov approximation. These approximations can be used for very 
low temperatures $T \ra 0$ and asymptotically weak interactions, and principally are not 
applicable for finite temperatures and sufficiently strong interactions. On the contrary,
the self-consistent approach can be employed for finite temperatures and strong interactions,
as has been shown for Bose-condensed systems in optical lattices \cite{Yukalov_56}, for
dipolar and spinor atoms \cite{Yukalov_57}, binary mixtures of components with BEC 
\cite{Rakhimov}, and for Bose-condensed systems in random potentials 
\cite{Yukalov_Graham,Yukalov_2007}. Not going into the details of the particular cases, which
can be found in the cited references, we may mention that the possibility of describing
finite temperatures and strong interactions results in the discovery of a number of interesting
effects not existing at weak interactions. For instance, the dipolar instability appearing 
in the spectrum of collective excitations is rather different from that arising under the 
Bogolubov approximation \cite{Yukalov_57}. The criterion of the phase transition between 
miscible and immiscible states in a binary mixture is essentially different at finite 
temperatures and strong interactions, as compared to the known condition of miscibility 
for zero temperature and weak interactions \cite{Rakhimov}. In the presence of finite 
interactions and strong disorder, in Bose-condensed systems there arises a Bose-glass 
fraction with peculiar properties \cite{Yukalov_Graham,Yukalov_2007}. What is the most 
important is the fact that the self-consistent approach opens the principal possibility 
of describing Bose-condensed systems at finite temperatures and strong interactions, which 
are not accessible in semi-classical and Bogolubov approximations, as well as using not 
self-consistent approximations. The latter, for instance, cannot correctly describe the 
Bose-Einstein condensation transition that in such approximations becomes of first order, 
while in the self-consistent approach it was shown \cite{Yukalov_2014} to be of correct 
second order for particle interactions of any strength.

\section{Normal versus symmetry-broken averages}

The previous section proves that the chemical potentials $\mu_0$ and $\mu_1$,
responsible for the validity of two different conditions, generally, are also
different. The potential $\mu_0$ guarantees that the condensate density $\rho_0$
is a minimizer of the thermodynamic potential, while the chemical potential $\mu_1$
makes the spectrum gapless. As is explained in Sec. XI, two chemical potentials are
necessary to ensure the correct definition of two particle numbers for condensed
$(N_0)$ and non-condensed $(N_1)$ particles. And there are no general restrictions
that would require the equality of these two chemical potentials.

Recall that the origin of the Hohenberg-Martin dilemma \cite{Hohenberg_33} of conserving
versus gapless approaches, discussed in Sec.XI, is directly caused by the assumption of 
equality of the chemical potentials $\mu_0$ and $\mu_1$. Therefore, if one sets these 
chemical potentials equal, then one returns back to the Hohenberg-Martin dilemma getting 
either the breaking of thermodynamic relations or a gap in the spectrum of collective 
excitations, which in both these cases imply the system instability. Thus, if our aim is 
an accurate description of a stable equilibrium Bose-condensed system, then, generally, 
we have to set two chemical potentials that do not need to be equal.

One may ask whether there exists a situation when the chemical potentials $\mu_0$ and $\mu_1$
could be set equal. The answer is straightforwardly seen from expressions (\ref{138}). Really,
under asymptotically weak interactions, when $n_k$ and $\sigma_k$ are asymptotically small,
then both chemical potentials are approximately equal, $\mu_0 \approx \mu_1$. This is the case
of the Bogolubov approximation, when one has $\mu_0 = \mu_1 = \rho \Phi_0$. Hence, the chemical
potentials can be set equal under the validity of the Bogolubov approximation, that is, when
temperature is low and interactions are asymptotically weak. However, at finite temperatures 
and finite interactions two chemical potentials are compulsory. 

As is seen from Eq. (\ref{138}), the difference in the chemical potentials is caused by
the existence of the so-called anomalous, or symmetry-broken average $\sigma_k$. For a
normal, non-condensed system, where there are no condensed particles but all particles
are normal, so that $\sigma_k$ is zero, the values of the chemical potentials coincide.
However, as soon as the system is Bose-condensed, there appear two kinds of particles,
condensed, whose number is $N_0$, and non-condensed, whose number is $N_1$. Then, to
correctly define these particle numbers, two Lagrange multipliers are compulsory, as
is elucidated in Sec. XI.

Dealing with symmetry-broken (anomalous) averages brings some inconvenience because the
integral anomalous average
\be
\label{139}
\sgm_1 \; = \; \frac{1}{V} \sum_{k\neq 0} \sgm_k
\ee
diverges and requires a regularization. For example, one may use the dimensional
regularization \cite{Yukalov_24,Andersen_34}.

Sometimes, one assumes that the simplest way of treating the symmetry-broken (anomalous) 
averages is to make the unjustified trick by omitting them at al. This trick was 
suggested by Shono \cite{Shono_58}. In recent literature, this trick is ascribed to Popov 
and is named the ``Popov approximation". However, the Shono trick makes the system 
unstable, its compressibility divergent, and BEC phase transition becomes of first order 
\cite{Yukalov_35}. In addition, Popov has never suggested such an improper trick, which 
is easy to check reading his works \cite{Popov_59,Popov_60,Popov_61} that are usually 
cited in this respect. Popov has nothing to do with the unjustified trick of neglecting 
the symmetry-broken (anomalous) averages.

The appearance of the symmetry-broken (anomalous) averages is the result of the gauge
symmetry breaking accompanying BEC. The same gauge symmetry breaking is responsible
for the arising Bose condensate. That is, both the symmetry-broken (anomalous) averages
as well as the condensate are connected with the same event of gauge symmetry breaking.
If there exists the Bose condensate, then there necessarily exist symmetry-broken
(anomalous) averages. Omitting the latter evidently makes the theory not self-consistent.
Actually, the Bose condensate and the anomalous average are equally important BEC order
parameters. And either both have to be taken into account below $T_c$ or both neglected
above $T_c$.

A reasonable approximation assumes that among several quantities, one of them is much
smaller than all others, so that the smallest quantity can be neglected. However, by
direct calculations it is straightforward to show \cite{Yukalov_63} that, depending on
the considered temperature, $|\sigma_k|$ is either larger than or comparable to $n_k$,
hence cannot be neglected.

In order to be more specific, let us compare explicitly the integral anomalous average
(\ref{139}) and the normal average yielding the density of non-condensed particles
\be
\label{140}
\rho_1 \; = \; \frac{1}{V} \sum_{k\neq 0} n_k   .
\ee
It is convenient to use the dimensionless quantities
\be
\label{141}
n_0 \; = \; \frac{\rho_0}{\rho} \; , \qquad 
n_1 \; = \; \frac{\rho_1}{\rho} \; , \qquad  
\sgm \; = \; \frac{\sgm_1}{\rho} \; 
\ee
characterizing the condensate fraction $n_0$, the fraction of non-condensed particles
$n_1$, and the dimensionless anomalous density (\ref{139}). We also introduce the
dimensionless sound velocity
\be
\label{142}
s \; \equiv \; \frac{ mc}{\rho^{1/3}} \;   .
\ee

Keeping in mind the local interactions with the potential
\be
\label{143}
 \Phi(\br) \; = \; 4\pi \; \frac{a_s}{m} \; \dlt(\br) \;  ,
\ee
where $a_s$ is scattering length, we define the gas parameter
\be
\label{144}
\gm \; \equiv \; \rho^{1/3} \; a_s
\ee
describing the strength of particle interactions.

Following the self-consistent approach of Sec. XIII, and using the techniques presented
in detail in Refs. \cite{Yukalov_64,Yukalov_65,Yukalov_66,Yukalov_67}, we can explicitly
calculate all quantities of interest. For instance, let us  concentrate on low-temperature 
properties of a Bose-condensed system, where all quantum effects are the most pronounced. 
For the fractions of condensed and non-condensed particles at zero temperature, we get
\be
\label{145}
 n_0 \; = \; 1 -\; \frac{s^3}{3\pi^2} \; , \qquad 
n_1 \; = \; \frac{s^3}{3\pi^2} \;  ,
\ee
with the sound velocity defined by the equation
\be
\label{146}
s^2 \; = \; 4\pi \gm ( n_0 + \sgm) \;   .
\ee
The anomalous (symmetry-broken) average (\ref{139}) is calculated by employing the
dimensional regularization, that is exact at infinitesimally small interactions, and
analytically continuing it to finite interactions. This gives \cite{Yukalov_67}
\be
\label{147}
 \sgm \; = \; 
\frac{8}{\sqrt{\pi}} \; \gm^{3/2} \; 
\left[ n_0 + \frac{8}{\sqrt{\pi}} \; \gm ^{3/2} \; \sqrt{n_0} \; 
\right]^{1/2} \;  .
\ee

At weak interactions $(\gamma \ll 1)$, we find the expansions for the sound velocity
\be
\label{148}
s \; \simeq \; 
\sqrt{4\pi \gm} + \frac{16}{3} \; \gm^2 -\; \frac{64}{9\sqrt{\pi}} \; \gm^{7/2} \;   ,
\ee
particle fractions of condensed atoms
\be
\label{149}
 n_0 \; \simeq \; 1 -\;
 \frac{8}{3\sqrt{\pi}} \; \gm^{3/2} -\; \frac{64}{3\pi} \; \gm^3 - \; 
\frac{256}{9\pi^{3/2}} \; \gm^{9/2} \;   ,
\ee
non-condensed atoms
\be
\label{150}
 n_1 \; \simeq \; 
 \frac{8}{3\sqrt{\pi}} \; \gm^{3/2} + \frac{64}{3\pi} \; \gm^3 + 
\frac{256}{9\pi^{3/2}} \; \gm^{9/2} \;   ,
\ee
and for the anomalous average
\be
\label{151}
 \sgm \; \simeq \; 
 \frac{8}{\sqrt{\pi}} \; \gm^{3/2} + \frac{64}{3\pi} \; \gm^3 - \;  
\frac{1408}{9\pi^{3/2}} \; \gm^{9/2} \;    .
\ee
As is seen, the anomalous average $\sigma$ is three times larger than the normal average
$n_1$, hence in no way $\sigma$ can be neglected. It is principally wrong omitting a
quantity that is larger than what is retained.

It is useful to emphasize that for the ground-state energy, the expansion in powers of
the gas parameter reads as
\be
\label{152}
E \; \simeq \; 4 \pi \gm^3 \; \left( 1 + \frac{128}{15\sqrt{\pi}} \; \gm^{3/2} +
\frac{128}{9\pi}\; \gm^3 - \; \frac{2048}{9\pi^{3/2}} \; \gm^{9/2} \right) \;   .
\ee
The two first terms here coincide with the Lee-Huang-Yang \cite{Lee_79,Lee_80,Lee_81}
expression
\be
\label{153}
 E_{LHY} \; = \; 
4 \pi \gm^3 \; \left( 1 + \frac{128}{15\sqrt{\pi}} \; \gm^{3/2}\right) \;  .
\ee

Therefore it is absolutely wrong to say that the Lee-Huang-Yang terms provide
``beyond-mean field corrections", as far as all those terms, plus some others, are
contained in the frame of the mean-field HFB approximation.

The behavior of the condensate fraction $n_0$, normal fraction $n_1$, and the anomalous
average $\sigma$ for arbitrary gas parameters is shown in Fig. 1, from where it is clear
that the anomalous average is larger than the normal average, hence cannot be neglected.
In Fig. 2, the theoretical condensate fraction is compared with the numerical Monte Carlo
simulations \cite{Rossi_68}. Close agreement with the Monte Carlo numerical results
confirms the high accuracy of the self-consistent approach. Good agreement with Monte
Carlo results \cite{Dubois_69,Dubois_70} has also been obtained for the condensate fraction
of trapped atoms, and the effect of the trap-center condensate depletion above the gas
parameter $\gamma = 0.3$ has been explained \cite{Yukalov_71}.

\begin{figure}[ht]
\begin{center}
\includegraphics[width=9cm]{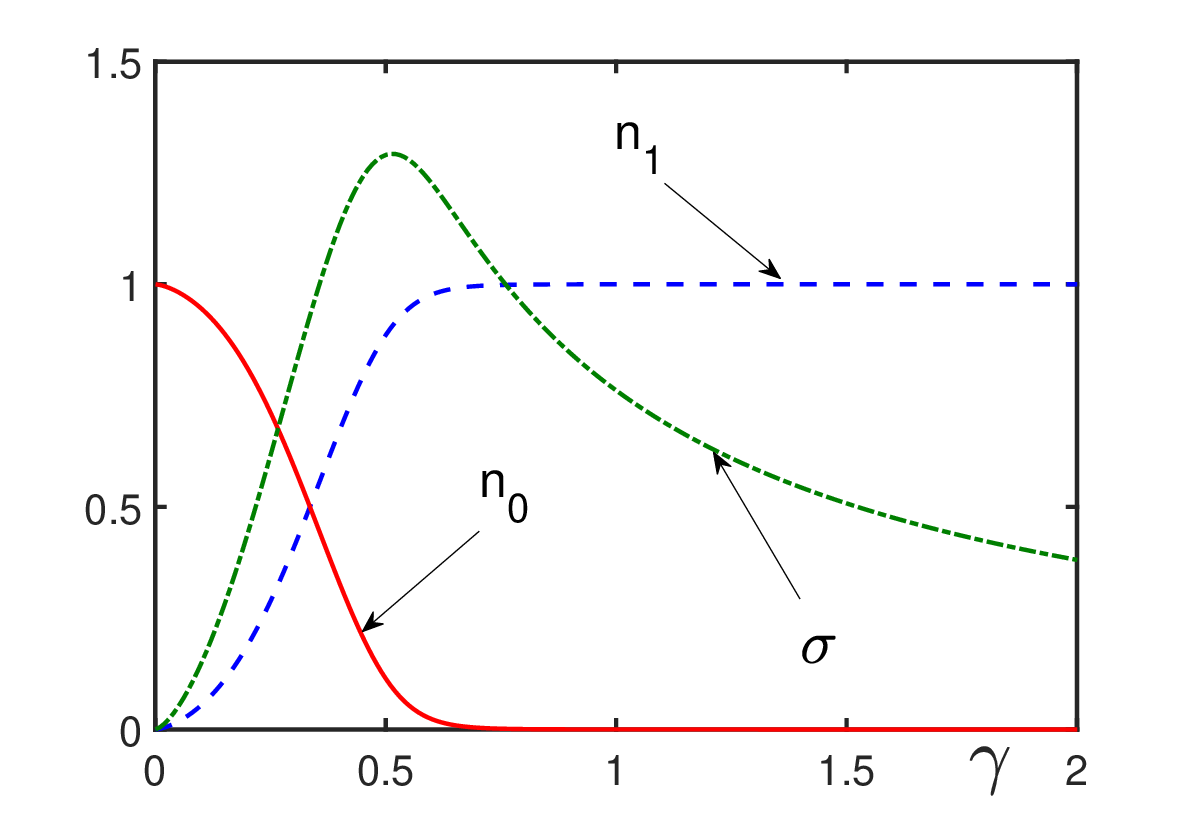}  
\end{center}
\caption{ \small
Condensate fraction $n_0$ (solid line), normal fraction $n_1$ (dashed line),
and anomalous average $\sgm$ (dashed-dotted line) as functions of the gas
parameter $\gm$.
}
\label{fig:Fig.1}
\end{figure}

\begin{figure}[ht]
\begin{center}
\includegraphics[width=9cm]{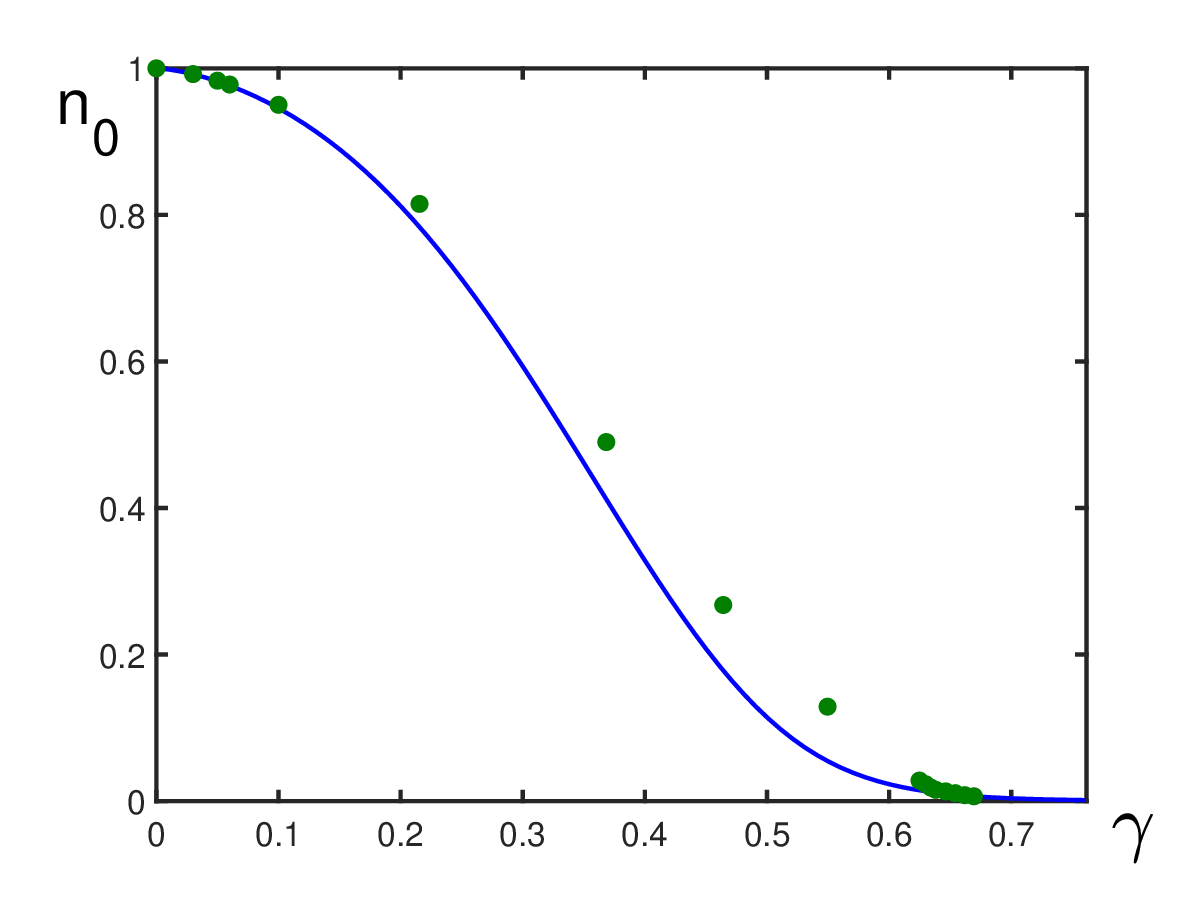}  
\end{center}
\caption{ \small
Condensate fraction $n_0$ (solid line), as a function of the gas
parameter $\gm$, compared with Monte Carlos calculations displayed by dots.
}
\label{fig:Fig.2}
\end{figure}

\section{Absence of thermodynamically anomalous fluctuations}

There are in literature wide spread claims on the occurrence in Bose-condensed systems,
in the whole range of their existence, of thermodynamically anomalous particle
fluctuations, exhibiting unphysical behavior with the variance
${\rm var} ({\hat N})$ proportional to $N^{4/3}$. They are named ``thermodynamically
anomalous", just ``anomalous" or ``non-thermodynamic" because, if such fluctuations
would really exist, then, according to the exact relations
\be
\label{154}
 \frac{{\rm var}(\hat N)}{N} \; = \; \rho \; T \varkappa_T \; = \; 
\frac{T}{ms^2} \; = \; S(0) \;  ,
\ee
the isothermal compressibility would diverge, sound velocity would become zero, and the
structure factor would be infinite, which, of course, is senseless implying the
instability of all equilibrium Bose-condensed systems, hence their overall non-existence.

The origin of this quandary comes from the following confusion. When calculating the
relative particle variance
\be
\label{155}
  \frac{{\rm var}(\hat N)}{N} \; = \; 
1 + \rho \int [\; g(\br) - 1 \; ] \; d\br \;  ,
\ee
where $g({\bf r})$ is a pair correlation function, one meets the divergent term
\be
\label{156}
 \frac{1}{\rho} \int \left[\; | \; \rho_1(\br,0) \; |^2 +
| \; \sgm_1(\br,0) \; |^2 \; \right] \; d\br \; = \;
\frac{1}{\rho} \int ( n_k^2 + \sgm_k^2 ) \; \frac{d\bk}{(2\pi)^3} \; = \; 
\frac{2}{\pi\rho\lbd_T^4} \; V^{1/3} \; ,
\ee
where $\lambda_T = \sqrt{2 \pi/m T}$. This divergence is exactly the same as that
happening for the ideal Bose gas in Sec. VIII. However there is a drastic difference
between these two cases. Calculations for the ideal Bose gas are exact, and the
compressibility divergence signifies the instability of the ideal-gas model. While for
the realistic Bose-condensed systems, such as trapped Bose gases and superfluid helium,
the occurring divergences are caused by calculational deficiencies.

The important point is the necessity of applying those approximations that do not distort
the nature of the model by changing its class. This is the principle of the model-class
preservation. The ideal Bose-gas model pertains to the Gaussian universality class, while
real Bose-condensed systems pertain to the class denoted as $XY$ or $O(2)$. When the
genuine model, with a Hamiltonian containing the product of four field operators, is
modified by resorting to an approximation reducing the initial Hamiltonian to a quadratic
Hamiltonian, one changes the model class by reducing the $O(2)$-class model to the
Gaussian-class model. Then in calculating some quantities there can appear the terms
induced by the Gaussian model, which are not related to the genuine model. The converting,
in the process of the approximation, of an $O(2)$-class model into a Gaussian-class model
is the reason for the appearance of the terms associated with the Gaussian model. These
spurious terms must be subtracted.

In more simple words, the same effect can be explained in purely mathematical language,
not plunging into physics. The considered model is of fourth order in field operators.
It is replaced by a second-order model with respect to field operators. This implies that
the quantities containing the products of field operators of the order higher than
second are not well defined and should be omitted. This directly concerns the calculation
of the particle fluctuations, as far as the variance of the number-of-particle operator
is expressed through the fourth-order terms. The application of a second-order
approximation to a fourth-order expression is not well defined, especially when the
involved approximation changes the nature of the model.

It is useful to notice that the same problem arises when an $O(2)$-class model is
approximated by any Gaussian-class model, whether using the HFB approximation, or just
the Bogolubov approximation, or a hydrodynamic approximation. In all the cases there appear
spurious Gaussian-type terms yielding the same divergence of compressibility. For all
these cases there is a simple and general cure: One has to subtract the spurious terms
caused by the approximation defect, limiting the consideration by the terms of the same
order as the used approximation. Employing the second-order approximation to expressions
of higher order requires to consider the terms up to second-order, subtracting the terms
of the order higher than the second.

When one resorts to some approximation, the general rule is not to distort the nature
of the given model. If, nevertheless, the approximation distorts the nature of the given
model , then the correct way is to supplement the approximation by the condition of the
subtraction of spurious terms induced by this distortion. In the case of approximating
a Bose-condensed $O(2)$ model by a Gaussian model, it is necessary to subtract the spurious
terms induced by the Gaussian approximation.

The explained above implies that, calculating the relative variance (\ref{154}), one has
to subtract the term (\ref{156}) as being beyond the region of the approximation
applicability, limiting the pair correlation function by the terms of second order,
which give
\be
\label{157}
g(\br) \; = \; 1 + \frac{2}{\rho} \; [\; \rho_1(\br,0) + \sgm_1(\br,0) \; ] \;  .
\ee
Then we have
\be
\label{158}
  \rho \int [\; g(\br) - 1 \; ] \; d\br \; = \; 
2\lim_{k\ra 0} ( n_k + \sgm_k) \; = \; \frac{T}{\Dlt} - 1 \;  ,
\ee
which results in the relative variance
\be
\label{159}
 \frac{{\rm var}(\hat N)}{N} \; = \; \frac{T}{m c^2} \;  .
\ee
This result shows that particle fluctuations in a Bose-condensed system are
thermodynamically normal \cite{Yukalov_24,Yukalov_32,Yukalov_43,Yukalov_64,Yukalov_72}.

Sometimes, one tries to connect the local value of the pair correlation function $g(0)$
with particle fluctuations, claiming that the thermodynamically anomalous fluctuations,
caused by the so-called ``grand canonical catastrophe" influence this value. However,
as is explained in Sec. VII, the occurrence of this ``catastrophe" is nothing but a mistake
due to the principal error of forgetting to break the gauge symmetry for a Bose-condensed
system.

The pair correlation function, according to Eq. (\ref{157}), can be written as
\be
\label{160}
g(\br) \; = \; 1 + \frac{2}{\rho} \int ( n_k + \sgm_k) \; e^{i\bk \cdot \br} \;
\frac{d\bk}{(2\pi)^3} \;   ,
\ee
so that its local value is
\be
\label{161}
 g(0) \; = \; 1 + \frac{2}{\rho} \; ( \rho_1 + \sgm_1) \; = \; 
1 + 2( n_1 + \sgm) \;  .
\ee
For weak interactions, using expansions (\ref{150}) and (\ref{151}), we obtain
\be
\label{162}
 g(0) \; \simeq \; 1 + \frac{64}{3\sqrt{\pi}} \; \gm^{3/2} +
\frac{256}{3\pi} \; \gm^3 \qquad ( \gm \ll  1) \;  .
\ee
Therefore, under weak interactions, when $\gamma \ra 0$, the pair correlation function
is close to one and is not able to demonstrate any ``grand canonical catastrophic regime"
that does not exist in principle.

Note that the Bose-condensed system of interacting particles in a trap, or in an
external potential exhibits \cite{Yukalov_78} the relative variance
$$
\frac{{\rm var}(\hat N)}{N} \; = \; 
\frac{T}{m} \int \frac{n(\br)}{c^2(\br)} \; d\br \;   ,
$$
where $n({\bf r}) = \rho ({\bf r})/N$ is the local fraction of particles and $c({\bf r})$
is the local sound velocity.

In that way, correct calculations demonstrate the absence of any thermodynamically
anomalous fluctuations in realistic equilibrium Bose-condensed systems. The appearance
of such unphysical thermodynamically anomalous fluctuations can be caused solely by
incorrect calculations.

\section{Fluctuation indices for composite observables}

The fluctuations of a physical observable, represented by an operator $\hat{A}$, can be
characterized by the fluctuation index \cite{Yukalov_74}
\be
\label{163}
\vp(\hat A) \; \equiv \; 
\lim_{N\ra\infty} \; \frac{\ln {\rm var}(\hat A)}{\ln N} \;   .
\ee
This index shows how the fluctuations grow with the increasing number of atoms, since
$$
{\rm var}(\hat A) \; \propto \; N^{\vp(\hat A)} \qquad ( N \ra \infty) \;    ,
$$
hence for extensive quantities
$$
\frac{{\rm var}(\hat A)}{\lgl \; \hat A\;\rgl} \; \propto \; N^{\vp(\hat A)-1} 
\qquad ( N \ra \infty) \;   .
$$
The fluctuations are thermodynamically normal, when $\varphi(\hat{A}) \leq 1$ and they
are thermodynamically anomalous, if $\varphi(\hat{A}) > 1$.

Sometimes, one states that, even if thermodynamically anomalous fluctuations of simple
observables cannot exist in stable equilibrium systems, but they could occur for composite
observables, for which thermodynamically anomalous fluctuations of some partial variances
could be compensated by mutual covariances. However, below it is explained that such a
compensation is impossible.

Let us consider a composite observable represented by an operator
\be
\label{164}
\hat A \; = \; \sum_i \hat A_i 
\ee
consisting of a sum of several linearly independent operators representing partial 
observables. The following theorem is proved \cite{Yukalov_32,Yukalov_64,Yukalov_72}.

\vskip 2mm
{\bf Theorem}. The fluctuation indices of the total operator (\ref{164}) and of its partial
sums are connected by the equality
\be
\label{165}
 \vp(\hat A) \; = \;  \sup_i \vp(\hat A_i) \;  .
\ee

\vskip 2mm
{\it Proof}. The variance of the operator (\ref{164}) reads as
\be
\label{166}
{\rm var}(\hat A) \; = \; \sum_i {\rm var}(\hat A_i) +
\sum_{i\neq j} {\rm cov}(\hat A_i,\; \hat A_j) \;  ,
\ee
where the covariance is
$$
 {\rm cov}(\hat A_i,\; \hat A_j) \; \equiv \; 
\frac{1}{2} \; \left( \hat A_i \; \hat A_j
+ \hat A_j \; \hat A_i \right) - 
\lgl \; \hat A_i \; \rgl  \lgl \; \hat A_j \; \rgl  \; .
$$
Employing the properties of quadratic forms \cite{Scharlau_75}, one has
\be
\label{167}
|\; {\rm cov}(\hat A_i,\; \hat A_j) \; |^2 \; \leq \;
{\rm var}(\hat A_i) \; {\rm var}(\hat A_j)  \;  .
\ee
Using this in Eq. (\ref{166}) and taking account of definition (\ref{163}) yields the
equality (\ref{165}).

\vskip 2mm

The direct conclusion from the above theorem is: {\it The fluctuations of a composite
observable are thermodynamically normal if and only if the fluctuations of all partial
observables are thermodynamically normal. The fluctuations of a composite observable
are thermodynamically anomalous if and only if at least one of the partial observables
exhibits thermodynamically anomalous fluctuations}.

\section{Equivalence of statistical ensembles}

Throughout the paper, we have used the second quantization representation in the frame 
of the grand canonical ensemble. The problem of ensemble equivalence is of great 
importance in physics \cite{Touchette_76}. In literature, it is possible to meet the 
statement that, for Bose-condensed systems, the grand canonical and canonical ensembles 
are not equivalent, since, although they lead to the coinciding averages for observables, 
but result in principally different particle-number fluctuations. The fluctuations of 
the total number-of-particle operator are different by definition, since in the grand 
canonical ensemble ${\rm var}({\hat N}) = {\rm var}({\hat N}_1)$, while in the canonical 
ensemble ${\rm var}({\hat N}) = 0$. And the condensate fluctuations in the grand canonical 
ensemble with broken gauge symmetry are absent at al., ${\rm var}({\hat N}_0) = 0$, 
while in the canonical ensemble they coincide with the fluctuations of non-condensed 
particles, ${\rm var}({\hat N}_0) = {\rm var}({\hat N}_1)$. This tells us that, 
strictly speaking, the condensate fluctuations separately are not well defined, which 
is reasonable, as far as condensed particles do not exist separately from non-condensed 
particles, but they always co-exist together in the frame of the same system. Hence one
cannot separate the fluctuations of condensed from non-condensed particles. In any case, 
one can assume that what is measured are the fluctuations of non-condensed particles.    

When somebody speaks on catastrophic condensate fluctuations in the grand canonical 
ensemble, the attentive reader can immediately understand that this statement is caused 
by the mistake connected to the so-called ``grand canonical catastrophe" that does not 
exist, as is explained in Sec. VII. This mistake arises when one compares not 
representative ensembles, while the meaning in a comparison exists solely for the 
comparison of representative ensembles \cite{Yukalov_35,Yukalov_64}.

Thus a grand canonical ensemble without gauge symmetry breaking cannot describe a
Bose-condensed system, as is explained in Sec. III. Then there is no any sense to
compare a not Bose-condensed system with a Bose-condensed one. Also, there is meaning
of comparing only the same observable quantities that can really be measured. For
particle fluctuations, the general quantity that can be measured and compared for any
ensemble is the relative variance of the number of uncondensed particles
${\rm var}({\hat N}_1)/N$. There are several publications comparing particle fluctuations
in different ensembles, the most accurate and complete of which is that accomplished by
Idziaszek \cite{Idziaszek_77}. He formally calculated the variances without subtracting
the Gaussian-induced terms. His main conclusions are as follows.

For the ideal uniform Bose gas, the microcanonical and canonical fluctuations of 
non-condensed particles have the relative variance scaling 
${\rm var}({\hat N}_1)/N \propto N^{1/3}$. The microcanonical and canonical fluctuations 
become equal in the thermodynamic limit.

For the ideal Bose gas in a harmonic potential, both, microcanonical and canonical
fluctuations are normal, scaling as ${\rm var}({\hat N}_1)/N \propto const$, but with
slightly different prefactors in the thermodynamic limit.

Interacting Bose gas, either uniform or in a harmonic potential, demonstrates canonical
and microcanonical fluctuations with the scaling ${\rm var}({\hat N}_1)/N \propto N^{1/3}$.
In the thermodynamic limit, the canonical and microcanonical variances always coincide.

The particle scaling for non-condensed particles, found in the paper by Idziaszek 
\cite{Idziaszek_77} for the microcanonical and canonical ensembles in three dimensions 
is in a complete agreement with the scaling presented in this paper for the grand 
canonical ensemble. This concerns Eq. (\ref{155}) as well as the Gaussian-induced term 
(\ref{156}). When the latter is subtracted, all particle fluctuations are normal and 
scaled identically. The ideal uniform Bose gas remains unstable, exhibiting 
thermodynamically anomalous fluctuations. And the interacting Bose gas, uniform as well 
as in external potentials, enjoys thermodynamically normal particle fluctuations.

In this way, if calculations are accomplished correctly, the fluctuations of non-condensed
particles, characterized by the relative variance ${\rm var}({\hat N}_1)/N$, are scaled, 
with respect to the number of particles $N$, identically in all ensembles. Slight 
differences in the prefactors can be caused by slightly different methods of their 
calculations. Therefore all statistical ensembles, microcanonical, canonical, and grand 
canonical, are equivalent with respect to the fluctuations of non-condensed particles.

\section{Conclusions}

As follows from the current literature on Bose-Einstein condensation, several major
points in the theory are often confused, which leads to incorrect conclusions and
principally wrong interpretation of experiments. These major issues are discussed and
explained in the present review. The following facts are elucidated.

\begin{itemize}
\item
Global gauge symmetry breaking is the necessary and sufficient condition for Bose-Einstein
condensation.

\item
A Bose-condensed and normal systems are described in different Fock spaces orthogonal
to each other in the thermodynamic limit.

\item
A thermodynamically stable statistical system enjoys a finite positive isothermal
compressibility.

\item
The isothermal compressibility is finite and positive then and only then when the
relative particle variance is finite.

\item
The so-called ``grand canonical catastrophe" for Bose-condensed systems does not exist,
being caused by the incorrect use of the grand canonical ensemble without gauge symmetry
breaking.

\item
In the grand canonical ensemble, condensate particle fluctuations are absent.

\item
Ideal uniform Bose gas in three dimensions is unstable, becoming stable for spatial
dimension $d > 4$.

\item
Ideal trapped Bose gas in a power-law trap is stable only for the confining dimension
$D > 2$.

\item
The resolution of the Hohenberg-Martin dilemma of conserving versus gapless theories
requires the use of a representative statistical ensemble.

\item
The condensate wave function is the vacuum field of the Fock space with broken gauge
symmetry.

\item
The omission of symmetry-broken anomalous averages for Bose-condensed systems is
principally wrong leading to inconsistences and divergences.

\item
Thermodynamically anomalous particle fluctuations for interacting Bose-condensed systems, 
arising due to the distortion of the $O(2)$-class model to the Gaussian-class model, must 
be subtracted.

\item
Repulsive particle interactions stabilize the uniform Bose-condensed system independently
of the interaction strength.

\item
Fluctuations of a composite observable, being the sum of linearly independent observables 
are thermodynamically normal if and only if all partial fluctuations are thermodynamically 
normal, and are thermodynamically anomalous if and only if at least one partial observable 
has thermodynamically anomalous fluctuations.

\item
All statistical ensembles, grand canonical, canonical, and microcanonical are equivalent
with respect to the scaling of fluctuations of non-condensed particles.

\item
Thermodynamically anomalous particle fluctuations, scaling as 
${\rm var}({\hat N}_1)/N \propto N^{1/3}$, being absent in a stable equilibrium system
with a large number of particles $N \gg 1$, cannot be observed in experiments with such 
systems.
\end{itemize}

\section*{Acknowledgement}

The author is grateful to V.S. Bagnato for numerous useful discussions and to E.P. Yukalova
for numerical analysis and advice.  

\section*{Conflict of Interest Statement}

The author has no conflicts to disclose.

\section*{Data Availability Statement}

Data sharing is not applicable to this article as no new data were created or analyzed
in this study.

\newpage

\end{document}